\documentclass[pof,showkeys,review]{revtex4-1}

\usepackage[utf8]{inputenc}
\usepackage{fullpage}
\usepackage{amsmath}
\usepackage{amsfonts}
\usepackage{amssymb}
\usepackage{graphicx}
\usepackage[font={small}]{caption}
\usepackage{subcaption}
\usepackage{subfig}
\usepackage{url}

\usepackage{color}
\usepackage{calc}
\usepackage[english]{babel}
\usepackage[T1]{fontenc}
\usepackage{graphicx}
\usepackage{verbatim}
\usepackage{enumerate}

\def\Bmp#1{ \begin{minipage}{#1} }
\def\Bmpc#1{ \begin{minipage}[c]{#1} }
\def\Bmpt#1{ \begin{minipage}[t]{#1} }
\def\Bmpb#1{ \begin{minipage}[b]{#1} }
\def\Emp{ \end{minipage} }
\newcommand{\revt}[1]{{\color{black}#1}}

\def\C{{\mathcal{C}}}

\def\O{{\mathcal{O}}}

\def\B{{\mathcal{B}}}

\def\tf0{\tilde{\varphi}_{0}}

\def\RR{{\mathbb{R}}}

\def\n{{\bf n}}

\def\x{{\bf x}}

\def\u{{\bf u}}
\def\0{{\bf 0}}

\begin{document}

\title{Wake Effects on Drift in Two-Dimensional Inviscid Incompressible Flows}
\author{Sergei Melkoumian}
\affiliation{School of Computational Science and Engineering, \\
McMaster University  \\
Hamilton, Ontario L8S4K1, CANADA}
\author{Bartosz Protas}
\email[Corresponding author, Email:]{bprotas@mcmaster.ca}
\affiliation{Department of Mathematics \& Statistics, \\
McMaster University  \\
Hamilton, Ontario L8S4K1, CANADA }

\date{\today}

\begin{abstract}
  This investigation analyzes the effect of vortex wakes on the
  Lagrangian displacement of particles induced by the passage of an
  obstacle in a two-dimensional incompressible and inviscid fluid. In
  addition to the trajectories of individual particles, we also study
  their drift and the corresponding total drift areas in the F\"oppl
  and Kirchhoff potential flow models. Our findings, which are
  obtained numerically and in some regimes are also supported by
  asymptotic analysis, are compared to the wakeless potential flow
  which serves as a reference. We show that in the presence of the
  F\"oppl vortex wake some of the particles follow more complicated
  trajectories featuring a second loop. The appearance of an
  additional stagnation point in the F\"oppl flow is identified as a
  source of this effect. It is also demonstrated that, while the total
  drift area increases with the size of the wake for large vortex
  strengths, it is actually decreased for small circulation values. On
  the other hand, the Kirchhoff flow model is shown to have an
  unbounded total drift area. By providing a systematic account of the
  wake effects on the drift, the results of this study will allow for
  more accurate modeling of hydrodynamic stirring.
\end{abstract}


\maketitle


\section{Introduction}
\label{sec:intro}

When a body passes through an unbounded fluid, it induces a net
displacement of fluid particles. The difference between the initial
and final positions of a fluid particle is defined as the particle's
``drift'' \cite{childress2009}, and plays an important role in
characterization of the stirring occurring in multiphase flows
\cite{e03a} and due to swimming bodies \cite{tc10a}. Hereafter we will
exclusively focus on flows with velocity fields stationary in a
suitable steadily translating frame of reference, and will consider
flows symmetric with respect to the flow centerline. Analysis of drift
in time-dependent flows is more involved and some efforts in this
direction have been made using methods of chaotic dynamics
\cite{jtz93a,zjt94a}.

\begin{figure}
\centering
\includegraphics[width=0.7\textwidth]{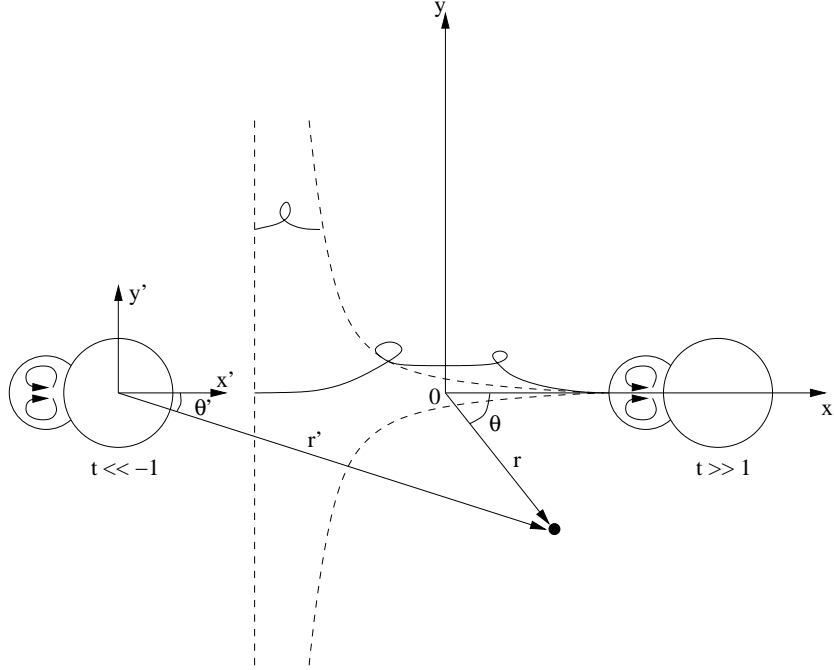}
\caption{Schematic of the problem indicating representative particle
  trajectories and the coordinate systems used.}
\label{fig:drift}
\end{figure}

Following the seminal study by Munk \cite{m66}, the phenomenon of
drift has recently received a lot of attention in the context of
mixing in the oceans caused by swimming organisms \cite{katija2009}.
However, most of the theoretical descriptions of stirring rely on
irrotational flow models used to compute or estimate the drift (an
exception to this is a recent study \cite{pushkin2013} focused on the
Stokesian approximation). The goal of the present contribution is to
understand the effect of vortex wakes on the drift in inviscid flows.
This will be accomplished in the two-dimensional (2D) setting using a
combination of careful numerical computations and mathematical
analysis. The set-up of the problem is illustrated in Figure
\ref{fig:drift} with $(r,\theta)$ and $(r',\theta')$ representing,
respectively, the polar coordinates in the fixed and moving frame of
reference.

Drift has been investigated for over a century with the earliest work
belonging to Maxwell \cite{maxwell1870} who showed that, when the
passing object is a circular cylinder inducing a simple potential
flow, then surrounding fluid particles follow trajectories in the form
of ``elastica'' curves (a more modern account of this problem can be
found in monograph \cite{mt1968}).  For a given fluid particle, an
elastica-shaped trajectory approaches a straight line parallel to the
path of the moving cylinder for points far upstream and downstream,
and exhibits a loop with a fore-and-aft symmetry (when the particle
travels along this loop, the cylinder is underneath it). The
historical origins and some other applications of elasticas are
surveyed in \cite{Levien:EECS-2008-103}.  Another major contribution
to this area is due to Darwin who, in addition to particle
trajectories, studied the problem of drift area and drift volume which
are global quantities characterizing the particle drift in a given
flow. Darwin's proposition \cite{darwin1953}, also referred to as a
``theorem'', is a key result relating the drift area or volume of a
moving body to its added mass.  Its utility consists in the fact that
the latter quantity tends to be easier to evaluate for flows past
objects with complex shape. There has been some debate
\cite{yih1985,benjamin1986,yih1997,ebh94a} concerning a rigorous proof
of this result in its full generality which was centered on the
evaluation method for conditionally convergent integrals. \revt{The
  relation between drift volume and added mass was investigated in a
  controlled experiment \cite{Bataille1991} where it was found that
  the shape of the displaced material surface is similar to that of
  the inviscid case and that the added mass coefficient measured for a
  spherical bubble for Reynolds numbers ranging from $Re = 500$ to
  $1000$ is consistent with its value obtained from Darwin's theorem}.
Connections between the Darwinian drift and the Stokesian drift,
related to the wave motion, were explored in \cite{em99a}.

The concept of drift was recently generalized for the case of flows
induced by propagating vortices (vortex rings) in \cite{d06}.
Motivated by biofluid applications, \revt{recent studies
  \cite{ltc11,pushkin2013}} investigated the effects of vortex wakes
on the drift induced by simple swimmers moving in the Stokes fluid.
\revt{On the other hand, recognizing that the concept of drift is
  idealized, in the sense that the object is assumed to travel during
  infinite time, corrections resulting from finite travel times were
  obtained leading to the definition of {\em partial drift}
  \cite{ebh94a,cmmv08}. This quantity was studied in recent
  experimental investigations concerning stirring by swimmers
  \cite{d06, katija2009}. Another related quantity is the mean squared
  displacement of particles which can be used to compute the effective
  diffusivity \cite{tc10a}.}  In the present study we provide a
thorough account of the effects of different vortex wakes on the drift
in inviscid flows. We will focus on 2D flows, because they offer
simple solutions amenable to straightforward analysis, so that
closed-form results can be obtained.

In our paper we begin in Section \ref{sec:drift} by precisely defining
the drift and the total drift area, and explaining how these
quantities can be evaluated in a given flow. Then, in Section
\ref{sec:flows}, we introduce the different vortex flows considered in
our study and identify their key parameters. Computational results are
presented in Section \ref{sec:results} together with a validation of
the numerical approaches, whereas their discussion and a posteriori
justification via asymptotic analysis are offered in Section
\ref{sec:analysis}.  Conclusions and outlook are deferred to Section
\ref{sec:final}.

\section{Drift: Definition and Calculation}
\label{sec:drift}

We will consider a circular cylinder of unit radius ($a = 1$) passing
through an incompressible inviscid fluid \revt{of unit density} in a
2D unbounded domain $\Omega$. \revt{We assume that the induced flow is
  potential and} the cylinder passes with its center along the
$x$-axis from $x = -\infty$ to $x = \infty$ with constant unit speed.
Hence, in the cylinder's frame of reference there is a uniform stream
at infinity such that $\u \rightarrow U \hat{\x}$ as $|\x| \rightarrow
\infty$, where $U = -1$ and $\hat{\x}$ is the unit vector associated
with the $x$-axis. In this frame of reference, the flow is steady,
i.e., $\u = \u(\x)$, and satisfies the no through-flow boundary
condition $\textbf{u} \cdot \textbf{n} = 0$ on the cylinder boundary
$\partial\Omega$, where $\n$ is the unit normal vector.  Euler system
\revt{describing the flow} is known to admit nonunique solutions and
different such solutions will be discussed below.

Hereafter, without the risk of confusion, we will interchangeably use
the vector and complex notation for various vector quantities. A point
$(x,y) \in \Omega$ will be represented as $\x = [x, y]^T$ or $z =
x+iy$, where $i := \sqrt{-1}$ (the symbol ``$:=$'' defines the
quantity on the left-hand side with the quantity on the right-hand
side). The fluid velocity will be denoted $\u(\x) = [u_x, u_y]^T$ or $V(z)
= (u_x-iu_y)(z)$, where $u_x$ and $u_y$ are the $x$ and $y$ components.
Assuming that the velocity field is incompressible and irrotational,
it will be expressed in terms of the complex potential $W(z) = (\phi +
i\psi)(z)$ as $V(z) = dW / dz$, where $\phi$ and $\psi$ are,
respectively, the scalar potential and the streamfunction.

In much the same way that Maxwell \cite{maxwell1870} and Darwin
\cite{darwin1953} studied the problem of drift, we will consider the
trajectories and drifts of individual particles in the fluid as the
cylinder passes. Let the initial position of the particle at $t=0$ be
$\x_0$ and $[x(t;\x_0), y(t;\x_0)]^T$ denote the corresponding
particle trajectory. Then, the displacement, or {\em drift}, of the
particle initially at $\x_0$ is defined as
\begin{equation} 
\xi(\x_0) := \int_{-\infty}^{\infty} u_x(x(t;\x_0), y(t;\x_0))\,dt,
\label{eq:xi}
\end{equation}
where the horizontal velocity component $u_x$ is given in the absolute
frame of reference. Integral \eqref{eq:xi} is improper and the
question of its convergence will be addressed further below. By
changing the integration variable from time $t$ to the polar angle
$\theta'$ in the moving frame of reference, cf.~Figure
\ref{fig:drift}, it can be transformed to an integral (still improper)
defined over a finite interval $\theta' \in [0,\pi]$ with the bounds
corresponding to the position of the particle in front and behind the
obstacle. Rewriting the velocity in the polar coordinate system in the
moving frame of reference as $\u = u_{r} \hat{\textbf{r}}' +
u_{\theta} \hat{\boldsymbol{\theta}}'$, where $\{\hat{\textbf{r}}',
\hat{\boldsymbol{\theta}}'\}$ are the two unit vectors, \revt{we may
  reformulate \eqref{eq:xi} as}
\begin{equation} 
\xi(\x_0) := \int_0^\pi  r' \frac{u_x(r',\theta')}{u_{\theta}(r', \theta')} \, d{\theta'}.
\label{eq:xi2}
\end{equation}
Form \eqref{eq:xi2} is more convenient for some of the manipulations
we will need to perform when deriving the drift in wakeless flow
(Section \ref{sec:wakeless}).

\revt{Historically, one of the quantities of interest in practical
  applications has been} the {\em total drift area} $D$ representing
the integral displacement of particles initially located on a line
perpendicular to the path of the obstacle at an infinite upstream
distance (Figure \ref{fig:drift})
\begin{equation}
D :=  2\int_{0}^{\infty}\xi(y_{\infty}) \, dy_{\infty}  = \int_{-\infty}^{+\infty}\xi(\psi) \, d\psi,
\label{eq:D}
\end{equation}
where $y_{\infty}$ is the transverse coordinate of the particle's
position when $t \rightarrow -\infty$ (with a slight abuse of
notation, $\xi$ may be equivalently considered a function of $\x_0$,
$y_{\infty}$ or $\psi$). The two integrals in \eqref{eq:D} are equal,
because $\psi \rightarrow y_{\infty}$ as $x \rightarrow \infty$. The
total drift area $D$ involves two nested improper integrals (in
expressions \eqref{eq:xi} and \eqref{eq:D}). Whether this quantity is
actually well-defined has been the subject of a debate
\cite{yih1985,benjamin1986,yih1997,ebh94a} with the conclusion that
this is indeed the case, provided the order of integration is as used
here, i.e., first with respect to the streamwise coordinate and then
with respect to the transverse coordinate. On the other hand,
reversing the order of integration will result in a conditionally
convergent expression. 

There are two ways to evaluate the total drift area $D$. First, we can
use a suitably transformed definition of formula \eqref{eq:D} combined
with the particle displacement given in \eqref{eq:xi}. From the
practical point of view, the most convenient way to evaluate the
improper integral \eqref{eq:xi} is to set the particle positions
$\x_0$ at $t=0$ and then obtain the trajectories by integrating the
system
\begin{equation}
\frac{d \x(t)}{dt} = \u(\x(t)), \quad \x(0) = \x_0
\label{eq:dxdt}
\end{equation}
forward and backward in time, i.e., for $t \rightarrow \pm \infty$,
for different $\x_0$. Since the initial particle positions in formula
\eqref{eq:D} are given at infinity, they need to be transformed to
positions with finite streamwise locations, e.g., $\x_0 = [0, y_0]^T$.
Since for a particle on a given streamline, $\psi$ is constant and
equal to some $C$, we have
\begin{equation}
C = \psi(0,y_0) = \lim_{x \rightarrow \infty} \psi(x,y_{\infty}) = y_{\infty}. 
\label{eq:C}
\end{equation}
Defining $g(y_0) := \psi(0,y_0) = y_{\infty}$ as the map between the
$y$-coordinates of the particle at $x = 0$ and at $x = \infty$, we
obtain
\begin{equation}
\frac{d y_{\infty}}{d y_0} = g'(y_0),
\label{eq:dg}
\end{equation}
so that \eqref{eq:D} becomes
\begin{equation}
D = 2\int_{1}^{\infty}\xi(g(y_0)) g'(y_0) \, dy_0, 
\label{eq:D2}
\end{equation}
where the lower bound is now set to unity, because the particle on the
streamline with $\psi = 0$ has the coordinate $y_0 = 1$ at $x_0 = 0$.
We note that function $g(y_0)$ will be different for different
solutions of \revt{the Euler equations governing the flow problem}.

The second method to evaluate the total drift area is to use Darwin's
theorem \cite{darwin1953} which stipulates that $D = M$, where $M$ is
the added mass, and the fluid density is assumed equal to the unity.
For our problem, the added mass is given by a line integral over the
contour $\C$ which is the boundary of the largest region with closed
streamlines
\begin{equation}
M = \oint \limits_\C  \phi n_x \, ds,
\label{eq:M}
\end{equation}
where $n_x$ is the $x$-component of the unit normal vector. In
addition to the boundary of the obstacle, contour $\C$ also comprises
the boundary of the recirculation region, if it is present in the
flow. The reason is that, in obtaining relation \eqref{eq:M}, the
divergence theorem cannot be applied on regions where singularities
(point vortices) are present.

Alternatively, one can bypass evaluation of integral \eqref{eq:M} by
the application of Taylor's added mass theorem \cite{t28a}. If we
consider the union of our cylinder and the recirculation region as a
single ``body'' $\B$ in motion, this theorem allows us to compute the
added mass in terms of the singularities within this region. Suppose
that our composite body contains $N$ sources and sinks with locations
$z_i$ and strength $m_i$. In addition, it contains $M$ doublets (or
dipoles) with strength $\mu_j$ and continuously distributed sources
and sinks with area density defined by $\sigma$. Then for irrotational
flows, the generalized form of the added mass given in
\cite{landweber1956} is
\begin{equation} 
\label{eq:gen_taylor} 
A_{\alpha 1} + B_{\alpha 1} + i
  (A_{\alpha 2} + B_{\alpha 2}) = 2 \pi \rho \left[ \int \limits_\B
    \sigma_{\alpha} z \, dA + \sum_{i = 1} ^ N m_{i \alpha} z_i + \sum_{j = 1} ^ M \mu_{j \alpha}\right], \quad \alpha=1,2,
\end{equation}
where $A$ is the added mass tensor and $B$ \revt{is} a tensor representing the mass
of the displaced fluid per unit area of the body with entries given by
\begin{equation}
B_{\alpha \beta} = \rho \oint \limits_\C x_{\beta} n_{\alpha} \, ds, \quad \alpha, \beta = 1,2
\label{eq:Bab}
\end{equation}
in which $x_1=x$, $x_2=y$, $u_1 = u_x$, $u_2 = u_y$.  For our problem, formula \eqref{eq:gen_taylor}
simplifies quite significantly. In particular, since we are
considering rectilinear motion in the $x$-direction of a body symmetric with the OX axis, we need only consider the element of the added mass tensor with $\alpha, \beta = 1$ so we may take the real part of \eqref{eq:gen_taylor} and drop these indices. Further, as there are no
continuous sources or sinks and $\rho = 1$, we get for the added mass
(now writing $A = M$)
\begin{equation} 
M = 2 \pi \, \Re\left[ \sum_{i = 1} ^ N m_i z_i + \sum_{j = 1} ^ M \mu_j \right] - B.
\label{eq:M2}
\end{equation}
In addition, since $ds$ is a infinitesimal distance along the body, we
have $n_x ds = dy$. Thus, $B$ can be simplified and interpreted as the
area of the cylinder augmented by the area of the wake
\begin{equation}
B = \oint \limits_\C x n_x \, ds = \oint \limits_\C x(y)\, dy.
\label{eq:B}
\end{equation}
We remark that relation \eqref{eq:M2} can be interpreted as consisting
of two parts: a ``universal'' part represented by the first term
involving only the far-field expansion of the velocity field induced
by the obstacle together with its vortex system and a second part
characterizing the specific flow and represented by $B$. An analogous
decomposition of the total drift area was obtained in
\cite{pushkin2013} for a swimmer in the Stokes flow. While all three
approaches, involving definition formula \eqref{eq:D}, added-mass
relation \eqref{eq:M} and Taylor's theorem
\eqref{eq:M2}--\eqref{eq:B}, are equivalent as far as the evaluation
of the total drift area is concerned, the first one offers additional
insights in the form of the particle trajectories responsible for the
observed drift.

\section{Model Problems}
\label{sec:flows}

In this Section we describe the three model flows we will consider in
our study. In addition to the wakeless potential flow for which the
questions of drift are well understood and which will serve as a
reference, we will also investigate the F\"oppl and Kirchhoff flows
which will be shown to have quite different properties. These two
flows are often invoked as the possible inviscid limits of steady
viscous Navier-Stokes flows \cite{srsk98}. For simplicity, in all
three cases the cylinder radius and the free stream at infinity have
unit values, $a=1$ and $U=-1$.

\subsection{Wakeless Potential Flow}
\label{sec:wakeless}

In the frame of reference attached to the obstacle, this flow \revt{is}
defined by the complex potential
\begin{equation}
W(z) = - \left(z + \frac{1}{z}\right)
\label{eq:W1}
\end{equation}
which does not involve any parameters. The flow field exhibits no
separation and is characterized by symmetry with respect to both OX
and OY axes. \revt{The streamline pattern is illustrated in Figure
  \ref{fig:traj}a}.

\subsection{F\"oppl Flow}
\label{sec:foppl}

The F\"oppl vortex system \cite{f13} is a one-parameter family of
solutions constructed by superimposing a pair of opposite-sign
vortices with circulations $\Gamma > 0$ and $-\Gamma$ located
symmetrically at $z_1=x_1 + iy_1$, $y_1 > 0$, and $\overline{z}_1$,
where the overbar denotes complex conjugation, on the flow with
potential \eqref{eq:W1}. The resulting potential of the F\"oppl flow
is thus
\begin{equation} 
	W(z) = -\left(z + \frac{1}{z}\right) + \frac{\Gamma}{2\pi i}
	\log\left(\frac{z-z_1}{z-\frac{1}{\overline{z_1}}}\right) 
	- \frac{\Gamma}{2\pi i}\log\left(\frac{z-\overline{z}_1}{z-\frac{1}{z_1}} \right).
\label{eq:W2}
\end{equation}
The locus of equilibrium vortex locations, the so-called F\"oppl curve,
is described by the algebraic relation
\begin{equation} 
\label{eq:r0}
r_1 ^2 - 1 =  2 \, r_1 \, y_1,
\end{equation} 
where $r_1 := \sqrt{x_1^2 + y_1^2}$. The circulation of the vortices
is related to their position through
\begin{equation} 
	\Gamma = 2 \pi \frac{(r_1^2 - 1) (r_1^4 -1)}{r_1^5}.
\label{eq:G0}
\end{equation}
For a given circulation $\Gamma > 0$, the F\"oppl system is a limiting
solution (as the vortex area goes to zero) of a family of Euler flows
with finite-area vortex patches discovered by Elcrat et al.
\cite{efhm00} (see also \cite{p06a}). The F\"oppl system features a
closed recirculation region with size growing with $\Gamma$. As is
evident from \eqref{eq:W2}, in the limit $\Gamma \rightarrow 0$ the
wakeless potential flow from Section \ref{sec:wakeless} is recovered.
\revt{The streamline patterns of F\"oppl flows with three
  representative values of the circulation $\Gamma$ are illustrated in
  Figures \ref{fig:traj}b-d}. The F\"oppl system has been successfully
employed as a model in a number of studies concerning the stability
and control of separated wake flows \cite{ta97,la03,p04a,p06c,p07c}.

\subsection{Kirchhoff Flow}
\label{sec:kirchhoff}

The Kirchhoff flow is a manifestation of the free-streamline theory of
the 2D ideal flows \cite{lc07}.  It features an object with two free
streamlines in the upper and lower half-planes that separate the
external fluid from the region behind the object, called the cavity
region, where the velocity is zero and the pressure is constant.
Commonly, the object used for these types of flows is a flat plate,
however, for consistency with the wakeless potential and the F\"oppl
flows, we will consider here a 1st-order approximation of a circular
cylinder presented in \cite{brodetsky1923}. The Kirchhoff flow is
interesting as an inviscid model, because it features an infinite wake
and a finite drag.

We will first clarify the notation: variable $z$ denotes the physical
plane we are interested in, where the circular cylinder is of unit
radius centered at $(0,0)$ and the flow is moving from right to left,
whereas variable $Z$ refers to the physical plane as used in
\cite{brodetsky1923}, where the cylinder instead has a radius of
approximately $1.77$ and is centered at approximately $(1.38,0)$ with
flow going in the opposite direction. We can define a map to switch
between the two spaces
\begin{equation}
Z(z) := -1.770434824562303 \, \overline{z} + 1.377445608362303.
\end{equation}

The complex potential is defined as a modified Levi-Civita
transformation \cite{brodetsky1923}
\begin{equation}\label{eq:W3}
W(\tau) = - \frac{\left( \tau - \frac{1}{\tau} \right)^2} {4}
\end{equation}
where $\tau = \rho e ^ {i \sigma}$ and $0 \leq \rho \leq 1$, $-\pi/2
\leq \sigma \leq \pi/2$. Unlike the models described in Sections
\ref{sec:wakeless} and \ref{sec:foppl}, potential \eqref{eq:W3} is not
given in terms of the variable in the physical space and additional
transformations are needed, so that it can be evaluated at $z$ or $Z$.
An intermediate map $\zeta(\tau)$ may be used to connect the $\tau$
and $Z$ planes
\begin{equation} \label{eq:zeta2}
\zeta = \frac{dZ}{dW}
\end{equation}
and for a 1st-order approximation of a circular cylinder we have
\begin{equation} \label{eq:zeta1}
\zeta(\tau) = \frac{ 1 + \tau } { 1 - \tau } e ^ {- 0.9426 \tau + 0.0191 \tau ^ 3}.
\end{equation}
Then, using the chain rule, we may write 
\begin{equation} \label{eq:dZ_dtau}
\frac{dZ}{d \tau} = \frac{dZ}{dW} \frac{dW}{d \tau},
\end{equation}
where the first derivative factor is \eqref{eq:zeta1} and the second
can be derived from \eqref{eq:W3}. Thus, $Z(\tau)$ can be determined
up to a constant through the integration
\begin{align} \label{eq:Ztau1}
Z(\tau) 	&= \int_a^{\tau_0} \frac{dZ}{dW} \frac{dW}{d \tau'}\, d\tau' \nonumber \\
			&= -\frac{1}{2} \int_a^{\tau_0} \left( \frac{ 1 + \tau' } { 1 - \tau' } \right) \left(1 + \frac{1}{{\tau}'^2} \right) \left( \tau' - \frac{1}{\tau'} \right) e ^ {- 0.9426 \tau' + 0.0191 {\tau'} ^ 3} \, d\tau'
\end{align}
where $\tau_0$ is an arbitrary constant. Integral \eqref{eq:Ztau1}
does not lend itself to analytical treatment, however, a generalized
series expansion for the integrand was found up to
$\mathcal{O}(\tau^2)$ around $\tau = 0$, so that, after integration,
we obtain
\begin{equation} \label{eq:Ztau2}
Z(\tau) = -\frac{1}{2} \left(c_1 \tau^{-2} + c_2 \tau^{-1} + c_3 \log{\tau} + c_4 \tau + c_5 \tau^2  \right)   + Z_0,
\end{equation}
where $c_1 = 0.5$, $c_2 = 1.0574$, $c_3 = -0.55904738$, $c_4 =
-0.8828122332$, $c_5 = 0.1113906656$ and $Z_0$ is some constant.

>From \eqref{eq:zeta2} and \eqref{eq:zeta1}, we can now compute the
velocities in the $Z$-plane in terms of the $\tau$ variable
\begin{subequations}
\label{eq:V3}
\begin{align} 
u_x(\tau) &= \Re \left( \frac{1}{\zeta (\tau)} \right), \label{eq:V3a} \\
u_y(\tau) &= -\Im \left( \frac{1}{\zeta (\tau)} \right). \label{eq:V3b}
\end{align}
\end{subequations}
Since we are interested in the flow in the direction opposite to the
one in the $Z$-plane \cite{brodetsky1923}, we set $V(z) = (u_x -
iu_y)(z) = (-u_x - iu_y)(\tau)$. In order to be able to evaluate
velocities \eqref{eq:V3} at a given location $Z$ in the physical
space, we need to invert map \eqref{eq:Ztau2}, i.e., find $\tau =
Z^{-1}(z)$. This is done by applying Newton's method to
\begin{equation} \label{eq:Ftau}
F(\tau) = -\frac{1}{2} \left(c_1 \tau^{-2} + c_2 \tau^{-1} + c_3 \log{\tau} + c_4 \tau + c_5 \tau^2  \right) - Z = 0.
\end{equation}
Once $\tau$ is found, the velocity at the required location can be
computed using \eqref{eq:V3}. \revt{The streamlines of the Kirchhoff
  flow can be seen in Figure \ref{fig:traj}e}.



\section{Results} 
\label{sec:results}

In this Section we compare the trajectories of individual particles,
their drift and the corresponding total drift areas in the three flows
introduced in the previous section. While, as reviewed below in
Section \ref{sec:wakeless_drift}, these quantities can be determined
analytically in the wakeless potential flow, they have to be computed
numerically in the case of the F\"oppl and Kirchhoff flows, and the
computational techniques are described and validated in Section
\ref{sec:compute}. Finally, the main results are presented in Section
\ref{sec:comparison}.

\subsection{Lagrangian Trajectories and Drift in the Wakeless
  Potential Flow}
\label{sec:wakeless_drift}

These classical results, recalled here for completeness, were derived
by Maxwell \cite{maxwell1870} and were also surveyed in \cite{mt1968}.
A key relation which makes this problem analytically tractable allows
one to express the radial coordinate of the particle in the cylinder's
frame of reference $r'$ in terms of its azimuthal angle $\theta'$ with
the streamfunction $\psi$ used as a parameter 
\begin{equation} 
\label{eq:r}
	r'(\theta') = \frac{\psi + \sqrt{\psi^{2} + 4a^{2} \sin^{2}{\theta'}}} 
	{2 \sin{\theta'}}.
\end{equation}
Then, when expressed using the angle $\eta$ made by the tangent to the
particle trajectory at a given point and the OX axis as the dependent
variable and the arc-length $s$ as the independent variable, the
equation governing the particle trajectories is of the form
\begin{equation}
\frac{d\eta}{ds} = \frac{4}{a^2} \left(y - \frac{1}{2}\psi\right)
\label{eq:elastica}
\end{equation}
implying that the trajectories are examples of ``elasticas'', a family
of curves with a long history in mathematics
\cite{Levien:EECS-2008-103}. The quantity ${d\eta}/{ds}$ represents
the curvature of the trajectory and the connection with elasticas was
first recognized by Milne-\revt{Thomson} \cite{mt1968}. \revt{However,
  for our purposes, it is more convenient to work with equation
  \eqref{eq:xi2} where the independent variable is changed from $t$ to
  $\theta'$.} Then, combining the resulting equation with relation
\eqref{eq:r} and integrating we obtain
\begin{subequations}
\label{eq:xy1}
\begin{align}
	x(u) &= \frac{a}{k}\left[\left(1-\frac{1}{2}k^{2}\right) u - E(u)\right], \label{eq:x1} \\
	y(u) &= \frac{a}{k} \left[\frac{dk}{d\psi} + \textrm{dn}(u) \right], \label{eq:y1}
\end{align}
\end{subequations}
where $k := {2a}/{\sqrt{\psi^2 + 4a^2}}$ (in our case $a=1$),
\begin{align*}
	E(u)         &:= \int_{0}^{\theta' - \pi/2} \sqrt{1-k^{2}\sin^{2}{\theta}} \, d{\theta}, \\
	\textrm{dn}(u) &:= \sqrt{1-k^{2}\sin^{2}({\theta' - \pi/2})}
\end{align*}
which are, respectively, an incomplete elliptic integral of the second
type and a Jacobi elliptic function. The variable $u$ parameterizing
trajectories \eqref{eq:xy1} is defined as an incomplete elliptic
integral involving the polar angle $\theta'$ (Figure \ref{fig:drift})
\begin{equation}
	u := \int_{0}^{\theta' - \pi/2} {\frac{1}{\sqrt{1-k^{2}\sin^{2}{\theta}}} \, d{\theta}}.
\end{equation}
We note that the initial position of the particle is encoded in the
value of the streamfunction $\psi$ appearing in the expression for
$k$. The drift corresponding to $t \in (-\infty,\infty)$,
cf.~\eqref{eq:xi} is then obtained by taking the limit $\theta'
\rightarrow \pi / 2$ in \eqref{eq:x1} which yields, after setting
$a=1$ and noting \eqref{eq:C},
\begin{equation}
\xi_1(y_0) = \frac{2}{k} \left[ \left(1 - \frac{1}{2}k^{2} \right)K - E \right],
\label{eq:xi1}
\end{equation}
where 
\begin{equation*}
	K =\int_{0}^{\pi/2} 
	\frac{1} {\sqrt{1 - k^{2}{\sin^2{\theta'}}}} \, d{\theta'}, \quad
	E = \int_{0}^{\pi/2}{\sqrt{1 - k^{2}{\sin^{2}{\theta'}}}} \,d{\theta'} 
\end{equation*}
are the complete elliptic integrals of the first and second type.
Using Darwin's theorem, the total drift volume can then be shown to be
\begin{equation}
D_1 = \pi.
\label{eq:D1}
\end{equation}
These results will be illustrated in Section \ref{sec:comparison}.

\subsection{Numerical Computation of Particle Trajectories, Drift and
  Total Drift Area in the F\"oppl and Kirchhoff Flows}
\label{sec:compute}

\begin{figure} 
\centering
\mbox{
\begin{subfigure}[b]{0.47\textwidth}
  \includegraphics[width=\textwidth]{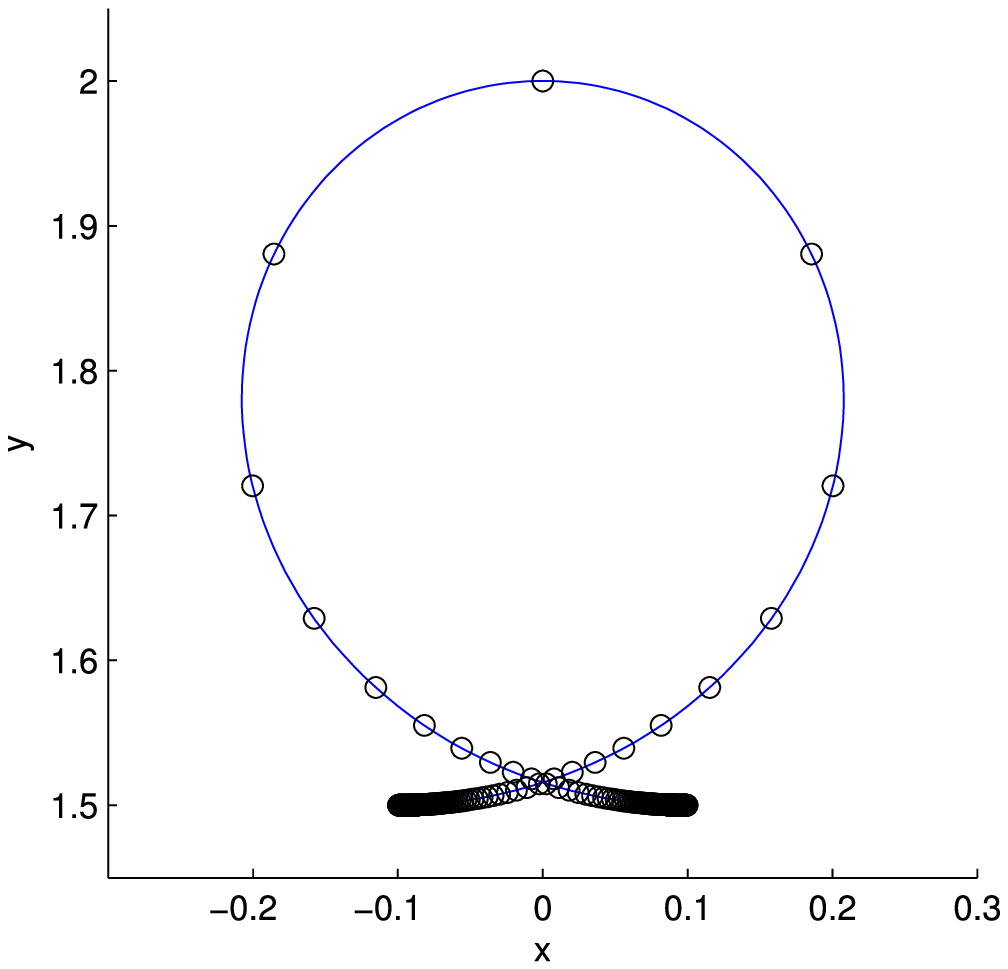}
  \caption{}
\end{subfigure}\quad
\begin{subfigure}[b]{0.52\textwidth}
  \includegraphics[width=\textwidth]{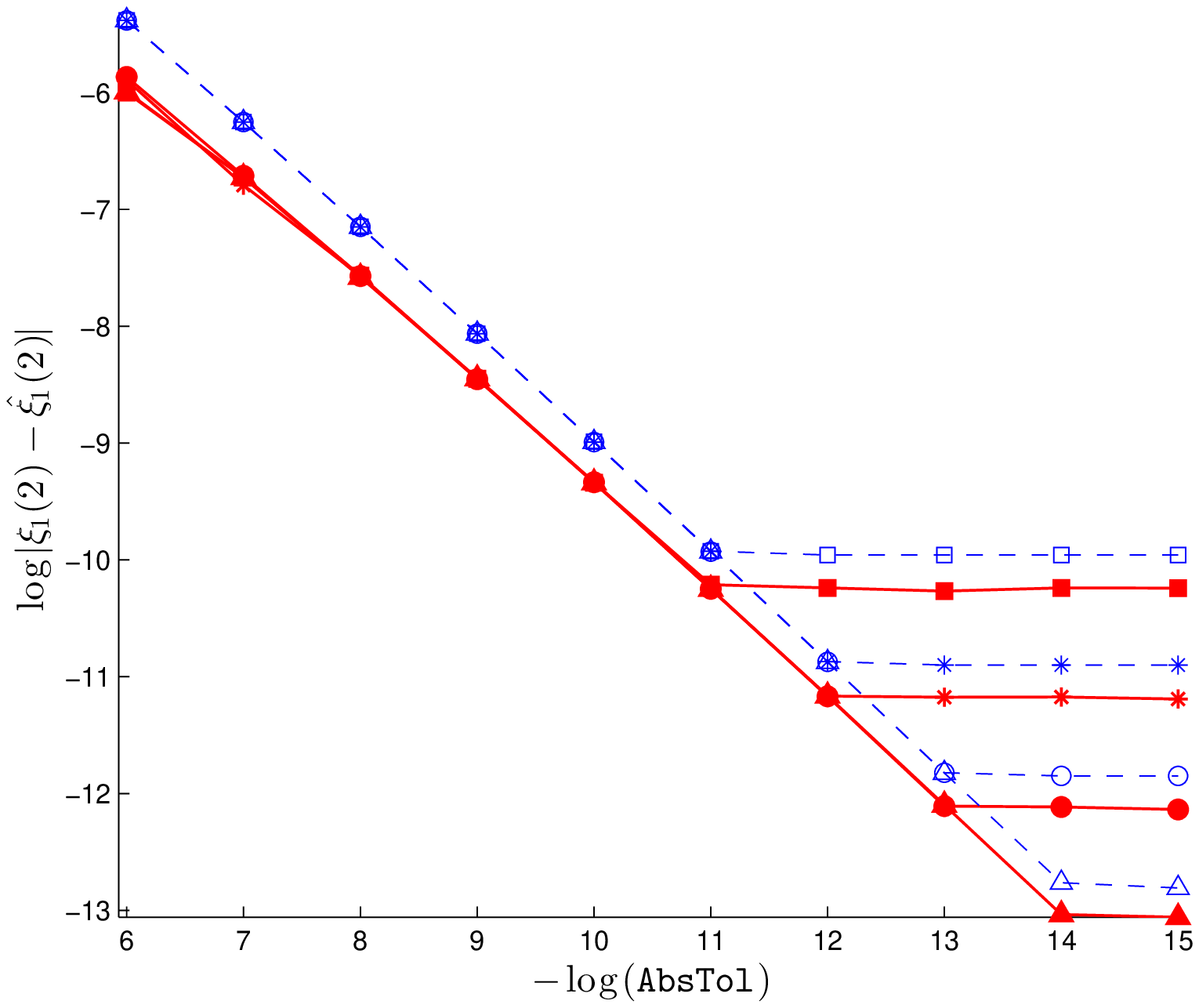}
  \caption{}
\end{subfigure}
}
\caption{(a) Particle trajectories corresponding to $y_0=2$ in the
  wakeless potential flow obtained numerically with $\texttt{RelTol} =
  \texttt{AbsTol} = 10^{-13}$ (symbols) and evaluated analytically
  using formulas \eqref{eq:xy1} (solid line); (b) error between the
  corresponding drift $\xi_1(2)$, cf.~\eqref{eq:xi1}, and its
  numerical approximation $\hat{\xi}_1(2)$ computed using routines
  \texttt{ode23} (open symbols) and \texttt{ode45} (filled symbols)
  for \texttt{RelTol} $= 10^{-10}$ (squares), $10^{-11}$ (stars),
  $10^{-12}$ (circles), and $10^{-13}$ (triangles).}
\label{fig:valid}
\end{figure}

Since explicit relations of the type \eqref{eq:r} are not available
for the F\"oppl and Kirchhoff flows, we need to resort to numerical
computations in order to determine the particle trajectories, drift
and the total drift area. The particle trajectories are computed as
described in Section \ref{sec:drift} by solving system \eqref{eq:dxdt}
with the initial data $\x_0 = [0, y_0]^T$, where $y_0 > 1$ is a
parameter (we note that, when $y_0 = 1$ in F\"oppl flow, the particle
is on the streamline connected to the stagnation point and the drift
$\xi(1)$ is infinite). In the case of the F\"oppl flow the particle
trajectories are additionally parameterized by the vortex circulation
$\Gamma$.  The velocity on the right-hand side of \eqref{eq:dxdt} is
obtained, respectively, by complex-differentiating potential
\eqref{eq:W2} and using expressions \eqref{eq:V3} in the two cases.
System \eqref{eq:dxdt} is integrated for different values of $y_0$
and, in the case of the F\"oppl flow, $\Gamma$ using MATLAB routines
{\tt ode23} and {\tt ode45} with adaptive adjustment of the time step.
Numerical evaluation of the drift, given by an improper integral
\eqref{eq:xi}, is a subtle issue requiring judicious choice of the
truncation $[-T,T]$ of the original unbounded interval
$(-\infty,\infty)$.  \revt{As shown in \cite{ebh94a,cmmv08}, such
  truncation of the integration domain leads to nontrivial corrections
  to the drift defined in \eqref{eq:xi} resulting in the so-called
  partial drift}. In order to exclude these finite-time effects from
the numerical integration, one has to make sure that $T$ is chosen
sufficiently large. For F\"oppl flow, this is achieved by setting $T$
close to \texttt{realmax}, the largest positive floating-point number
in the IEEE double-precision standard \cite{IEEEdp}, which is of the
order $\O(10^{300})$ and then balancing the accuracy with the
computational time by adjusting the relative and absolute tolerances,
{\tt RelTol} and {\tt AbsTol}, in the routines {\tt ode23} and {\tt
  ode45}. Owing to the adaptive adjustment of the time step employed
in these routines, the total computational time required for a single
particle trajectory does not typically exceed one minute on a
state-of-the-art workstation even for the finest tolerances. This
approach is validated by computing the particle trajectories
$\x(t;y_0)$ and the associated drift $\xi(y_0)$ numerically for the
wakeless potential flow (obtained setting $\Gamma = 0$ in
\eqref{eq:W2}) and then comparing them to the analytical expressions
\eqref{eq:xy1} and \eqref{eq:xi1} (since these formulas involve
special functions, care must be taken to enforce a required level of
precision in the evaluation of these functions as well). The results
obtained for a single trajectory with $y_0=2$ are presented in Figure
\ref{fig:valid}a, where we show a segment of the particle trajectory
computed numerically and given by expression \eqref{eq:xy1}, and in
Figure \ref{fig:valid}b in which we show the difference between the
exact drift value $\xi_1(2) = 2.011398641052742 \times 10^{-1}$ and
its numerical approximation $\hat{\xi}_1(2)$ for different fixed {\tt
  RelTol} and varying {\tt AbsTol}. As is evident from Figure
\ref{fig:valid}b, the error in the evaluation of the drift is rather
small and decreases algebraically with the refinement of both {\tt
  RelTol} and {\tt AbsTol}, Thus, in all subsequent calculations we
will use routine {\tt ode45} with $\texttt{RelTol} = \texttt{AbsTol} =
10^{-13}$.

Unlike in the case of the wakeless and F\"oppl flow where $T$ was
allowed to extend close to \texttt{realmax}, in Kirchhoff flow we have
to restrict the truncation of the time axis to $T = 10^3$ which is due
to the failure of Newton's method applied to \eqref{eq:Ftau} to
converge for such large values of $t$. However, since the structure of
the flow advecting the particles does not change much when $|t| > T$,
we will compensate for this by extrapolating the velocity for large
times.  Since, as will be shown below, the velocity field following
the particle trajectory is for sufficiently large $t$ a power-law
function of time, this extrapolation will be performed using the
formula
\begin{equation}
h(t) = ct^{\beta},
\label{eq:h}
\end{equation}
where $c \in \RR$ and $\beta < 0$, using 10 data points corresponding
to the largest available times.

As regards evaluation of the total drift area $D$, three different
approaches can be used: definition formula \eqref{eq:D}, or more
conveniently \eqref{eq:D2}, added-mass formula \eqref{eq:M} and
Taylor's theorem \eqref{eq:M2}--\eqref{eq:B}. In the first approach the
parameter space $y_0$ is discretized in such a way that the relative
variation of $\xi(y_0)$ between two adjacent discrete values of $y_0$
would not exceed $1\%$. The function $g(y_0)$ and its derivative
needed in \eqref{eq:D2} are identified for the F\"oppl flow as follows
\begin{align}
g(y_0) & = - \left( y_0 - \frac{1}{y_0} \right) + \frac{\Gamma}{2 \pi} \Bigg[ \log \left( \frac{\sqrt{x_1^2 + (y_0 + y_1)^2}}{\sqrt{x_1^2 + (y_0 - y_1)^2}} \right) \nonumber\\
& \hspace*{2.5cm}
+ \log \left( \frac{\sqrt{\left(\frac{x_1}{x_1^2 + y_1^2}\right)^2 + \left(y_0- \frac{y_1}{x_1^2 + y_1^2}\right)^2}}{\sqrt{\left(\frac{x_1}{x_1^2 + y_1^2}\right)^2 + \left(y_0 + \frac{y_1}{x_1^2 + y_1^2}\right)^2}} \right) \Bigg], 
\label{eq:g2} \\
g'(y_0) & = - \left( 1+\frac{1}{y_0^2} \right) + \frac{\Gamma}{2\pi} \Bigg[\frac{y_0 + y_1}{x_1^2 + (y_0 + y_1)^2} - \frac{y_0 - y_1}{x_1^2 + (y_0 - y_1)^2} \nonumber\\ 
& \hspace*{1.5cm}
+ \frac{y_0 - \frac{y_1}{x_1^2 + y_1^2}}{\left(\frac{x_1}{x_1^2 + y_1^2}\right)^2 + \left(y_0 - \frac{y_1}{x_1^2 + y_1^2}\right)^2} - \frac{y_0 + \frac{y_1}{x_1^2 + y_1^2}}{\left(\frac{x_1}{x_1^2 + y_1^2}\right)^2 + \left(y_0 + \frac{y_1}{x_1^2 + y_1^2}\right)^2} \Bigg].
\label{eq:dg2}
\end{align}
Concerning the computation of the total drift area via Taylor's
theorem \eqref{eq:M2}--\eqref{eq:B}, the two F\"oppl vortices and
their images inside the cylinder make the contributions $m = \pm
\frac{\Gamma}{2 \pi i}$ each, whereas the dipole at the origin
contributes $\mu = Ua^2$. Therefore, after setting $a=1$ and $U=-1$,
equation \eqref{eq:M2} becomes
\begin{equation}
\begin{aligned}
M & = 2 \pi \Re \left( \frac{\Gamma}{2 \pi i} z_1 -\frac{\Gamma}{2 \pi i \overline{z_1}}  -\frac{\Gamma}{2 \pi i} z_2 + \frac{\Gamma}{2 \pi i \overline{z_2}} -1 \right)  - B \\ 
  & = - 2 \pi  + 2 \Gamma \left( y_1 - \frac{ y_1}{x_1 ^2 + y_1 ^2} \right) - B.
\end{aligned}
\label{eq:MM}
\end{equation}
Since it does not appear possible to find an analytic expression for
$B$ representing the area of the recirculation bubble, it has to be
evaluated numerically using \eqref{eq:B}.

\subsection{Comparison of Particle Trajectories, Drift and Total Drift
  Area in Flows with Different Wake Models}
\label{sec:comparison}

The particle trajectories corresponding to several different initial
positions $\x_0 = [0, y_0]^T$ are shown in Figure \ref{fig:traj} for
the wakeless potential flow, the F\"oppl flow with different
circulations $\Gamma$ and for the Kirchhoff flow (the wakeless
potential flow is obtained as the F\"oppl flow with $\Gamma = 0 $).
The initial positions corresponding to the indicated cylinder
locations are marked with circles, whereas crosses indicate the
particle positions at the same instances of time in the different
cases. First, we see that in the wakeless potential flow (Figure
\ref{fig:traj}a) all the particle trajectories have the form of
elastica curves symmetric with respect to the OY axis. The presence of
the vortex wakes in the F\"oppl flows breaks the fore-and-aft symmetry
of the trajectories and, for small values of $y_0$, spawns a second
loop on the trajectory which becomes larger for increasing vortex
circulations $\Gamma$. The presence of this secondary loop results
from the fact that, for sufficiently small $y_0$, the transverse
component $u_y$ of the particle velocity must change sign when the
particle is flowing around the recirculation region. Supplementary
material, available on-line \cite{SuppMat}, contains animated versions
of particle trajectories {for a single particle close to the
  cylinder's line of motion and for particles that are initially
  aligned vertically as indicated in Figure \ref{fig:drift}}.

\begin{figure}
  \begin{subfigure}{\textwidth}
    \includegraphics[width=\textwidth]{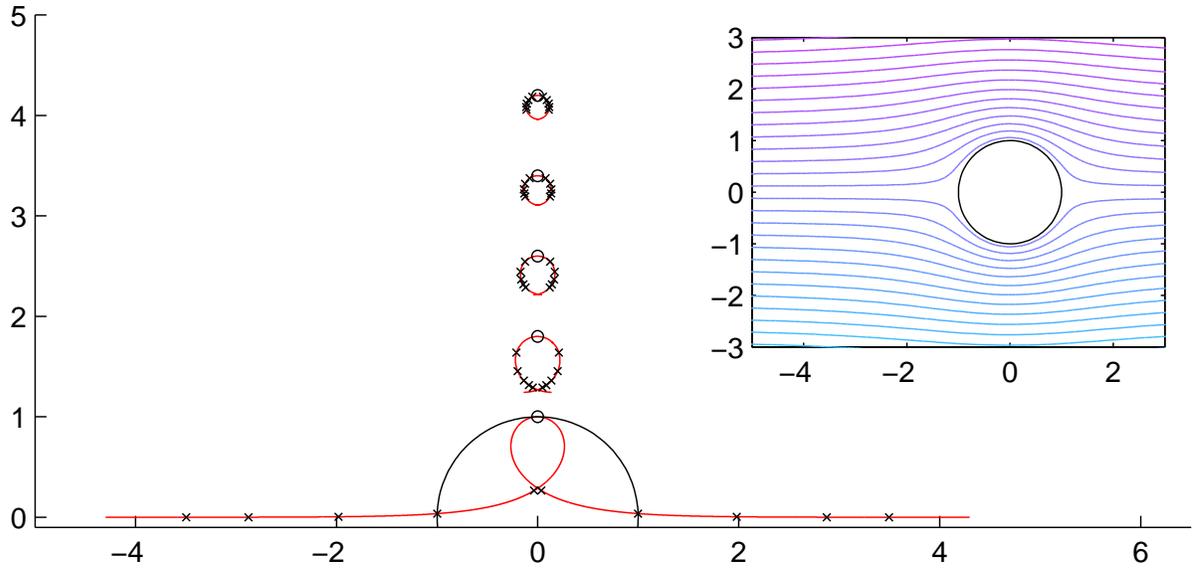} 
    \caption{wakeless potential flow ($\Gamma = 0$)} 
    \label{fig:stacked_gamma_0}
     \vspace*{1cm}
  \end{subfigure}
  \begin{subfigure}{\textwidth}
    \includegraphics[width=\textwidth]{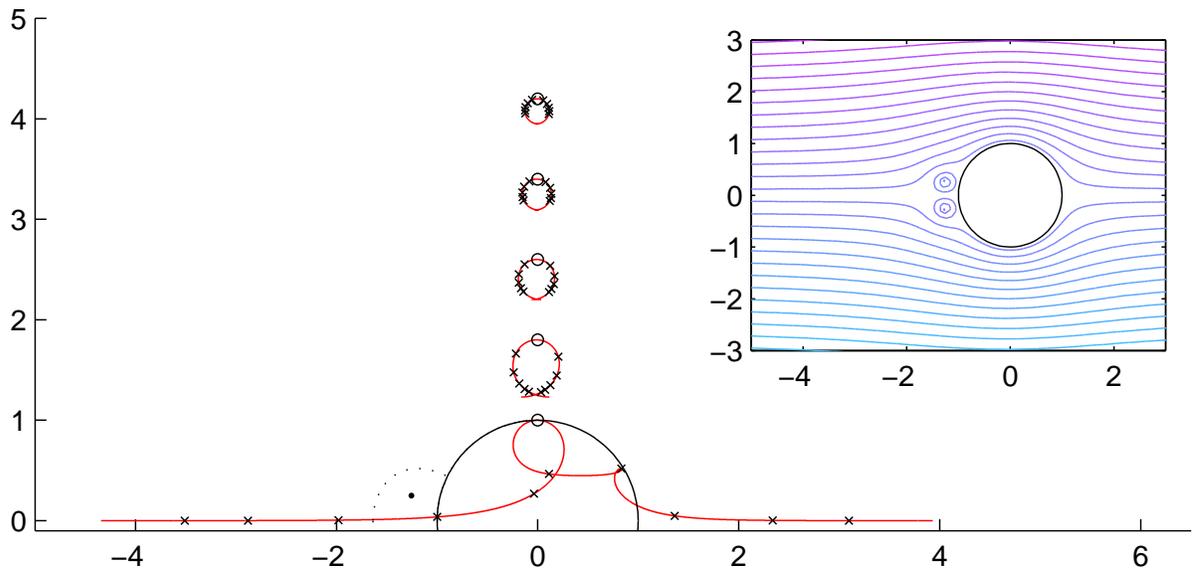}
    \caption{$\Gamma = 1.9663$}
    \label{fig:stacked_gamma_1}
  \end{subfigure}
  \caption{Particle trajectories for different initial conditions
    $\x_0=[0,y_0]^T$ in the wakeless potential flow (a), the F\"oppl
    flow with different circulations (b,c,d) and the Kirchhoff flow
    (e), \revt{with the insets illustrating the streamline patters of
      the flows.}  The x's represent the particle positions at unit
    time intervals, whereas the o's correspond to the particle
    positions at $t = 0$, at which the cylinder, recirculation bubble
    for F\"oppl flow and the cavity for Kirchhoff flow are also
    indicated . The total drift areas produced by the fluid
    displacements shown in figures (a) and (c) are approximately
    equal, cf.~\eqref{eq:D2}, even though the individual particles
    with the same initial locations have quite different
    trajectories.}
\label{fig:traj}
\end{figure}
\begin{figure}
\addtocounter{figure}{-1}
\setcounter{subfigure}{2}
  \begin{subfigure}{\textwidth}
    \includegraphics[width=\textwidth]{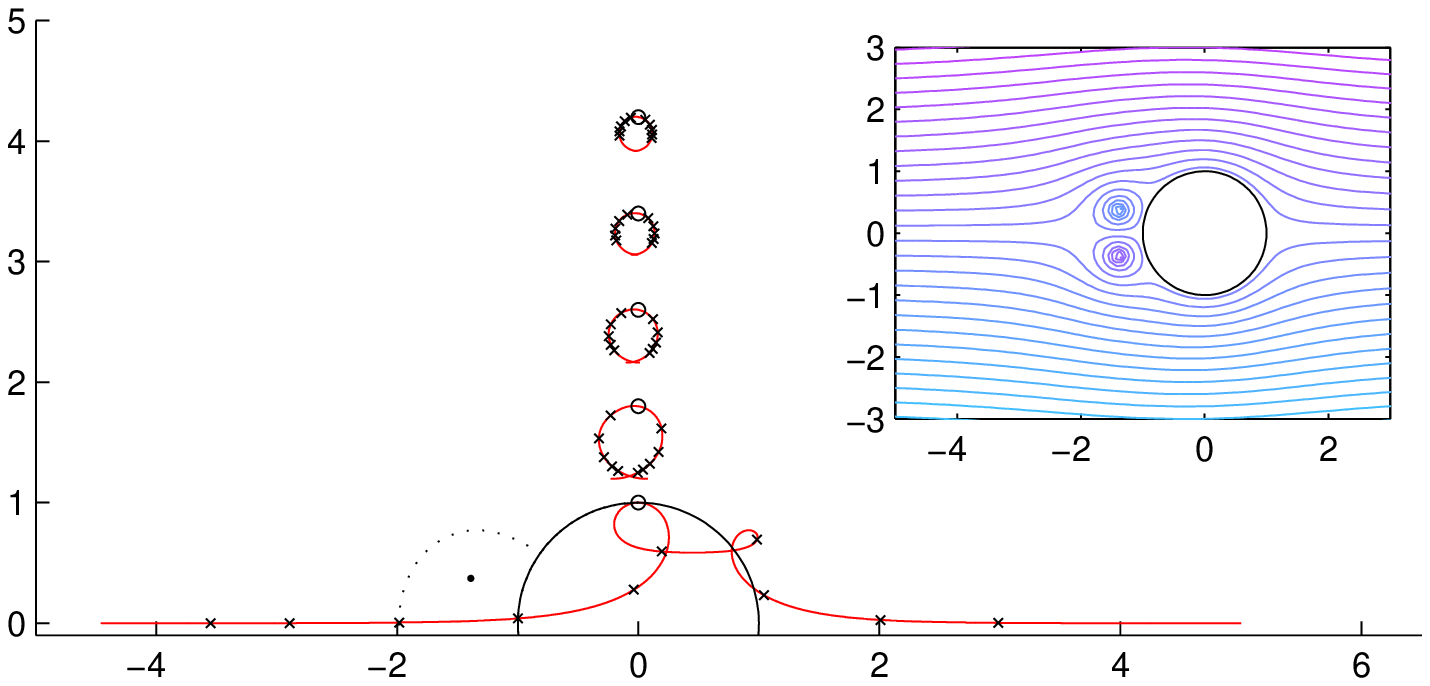}
    \caption{$\Gamma = 3.595$}
    \label{fig:stacked_gamma_3}
     \vspace*{1cm}
  \end{subfigure}
  \begin{subfigure}{\textwidth}
    \includegraphics[width=\textwidth]{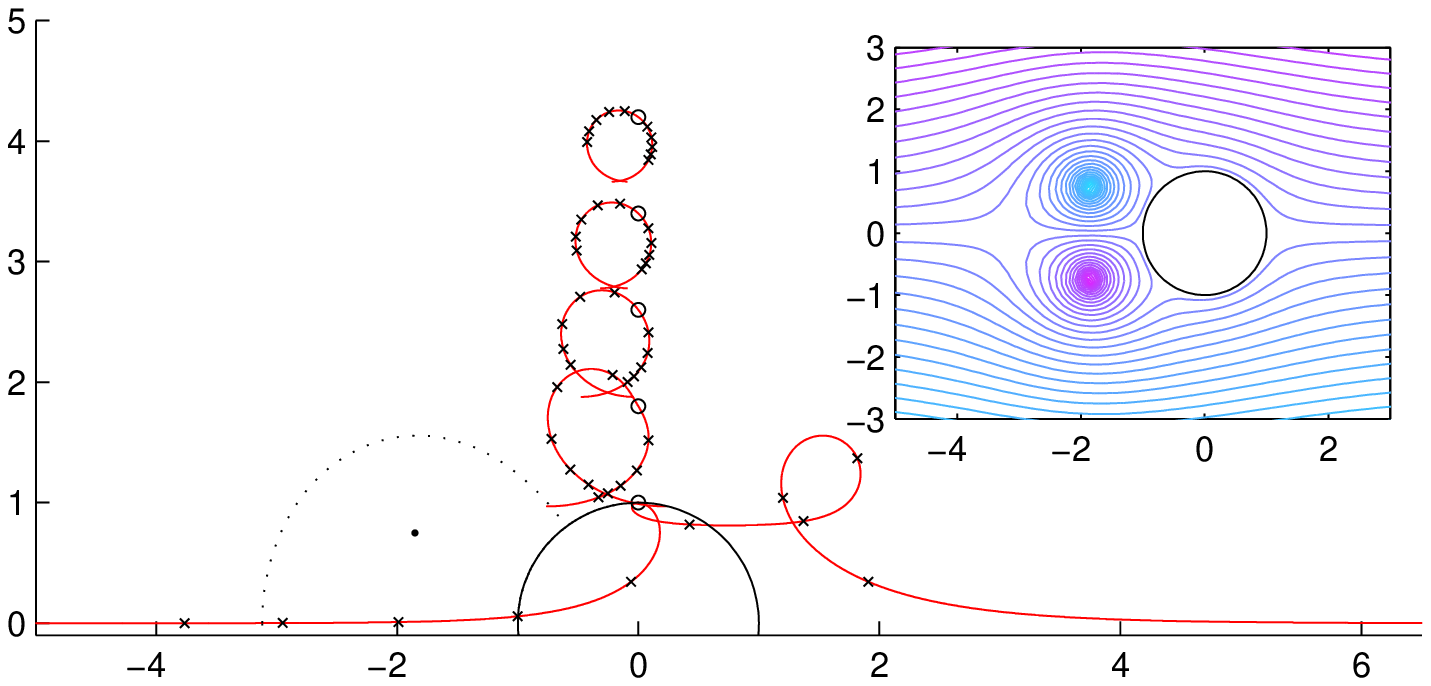}
    \caption{$\Gamma = 8.8357$} 
    \label{fig:stacked_gamma_4}
  \end{subfigure}
  \caption{(Continued, see previous caption for details)}
\end{figure}
\begin{figure}
\addtocounter{figure}{-1}
\setcounter{subfigure}{4}
    \begin{subfigure}{\textwidth}
    \includegraphics[width=\textwidth]{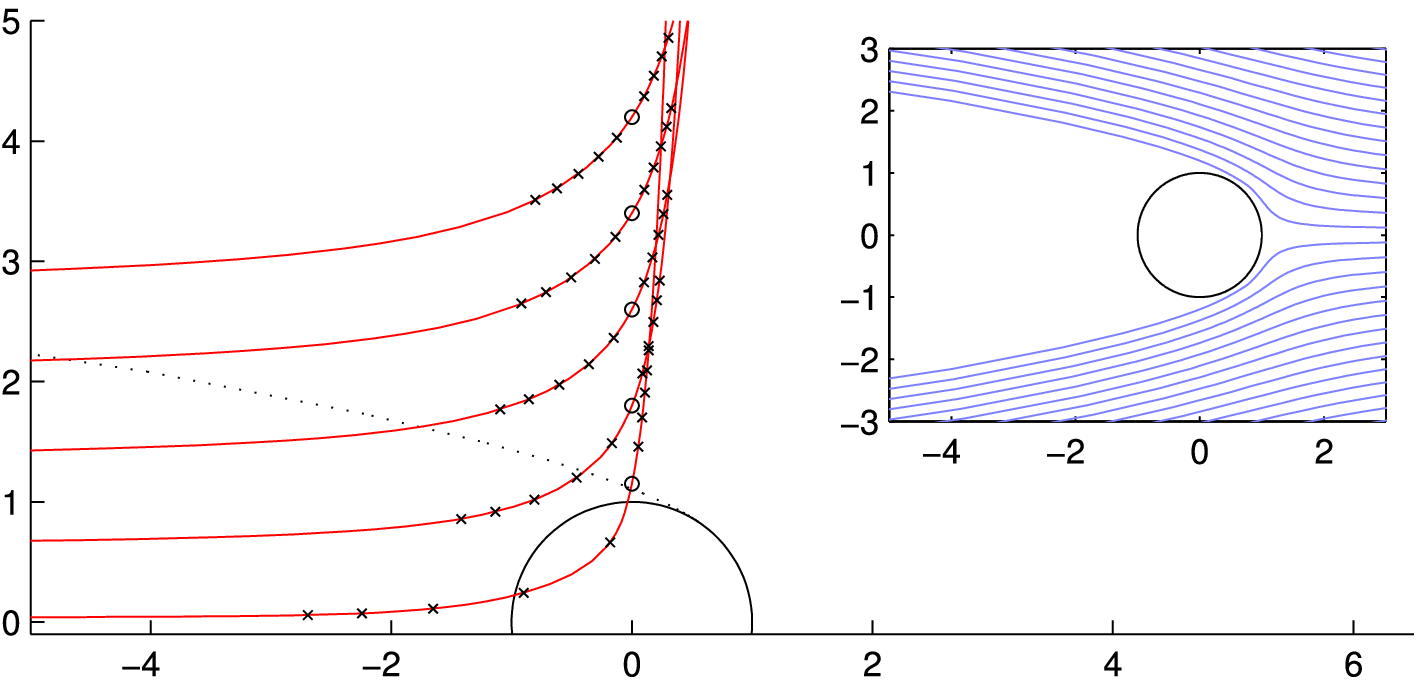}
    \caption{Kirchhoff flow} 
    \label{fig:stacked_kirchhoff}
  \end{subfigure}
  \caption{(Continued, see previous caption for details)}
\end{figure}

\begin{figure}
\centering
\mbox{
\begin{subfigure}[b]{0.5\textwidth}
  \includegraphics[width=\textwidth]{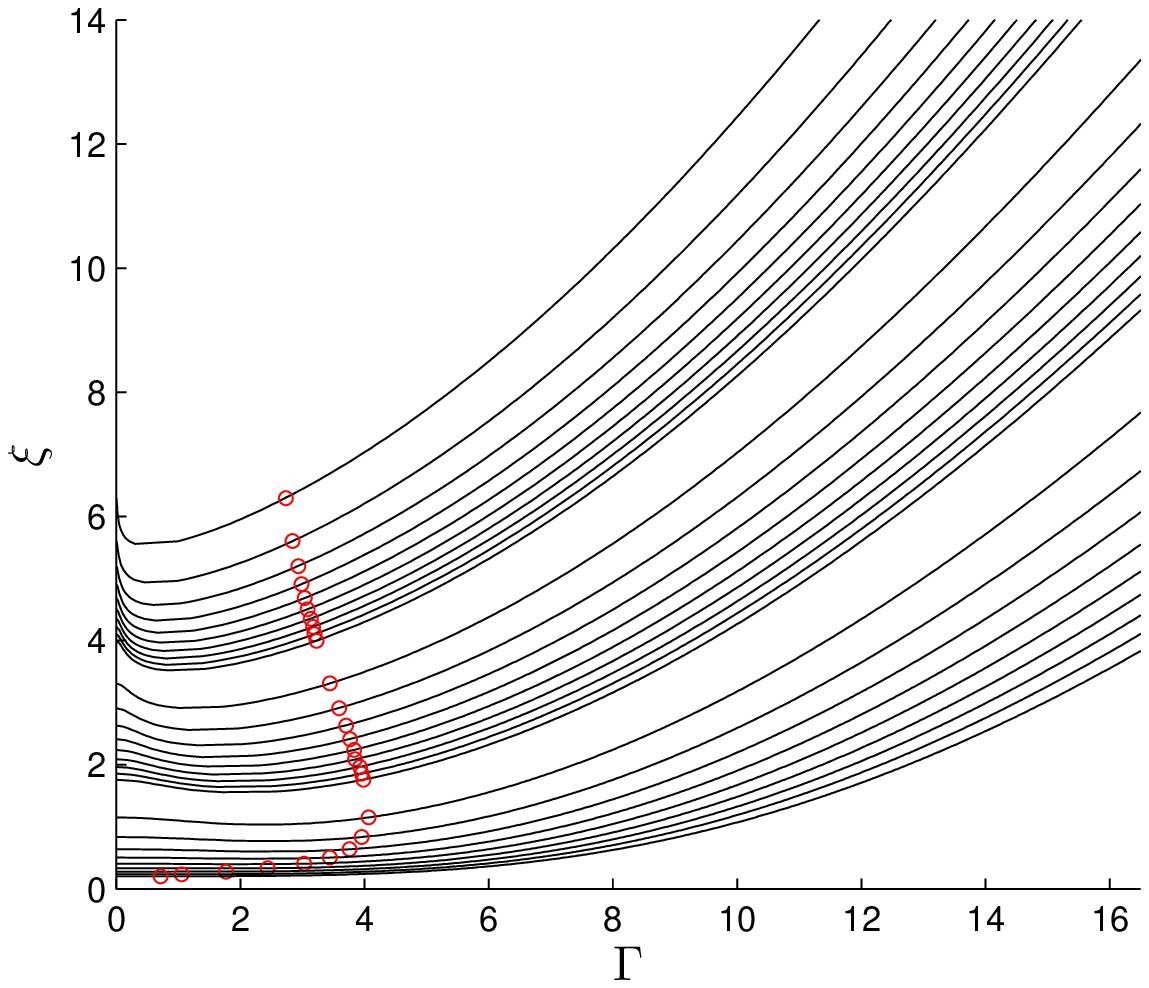}
  \caption{}
\end{subfigure}
\begin{subfigure}[b]{0.5\textwidth}
  \includegraphics[width=\textwidth]{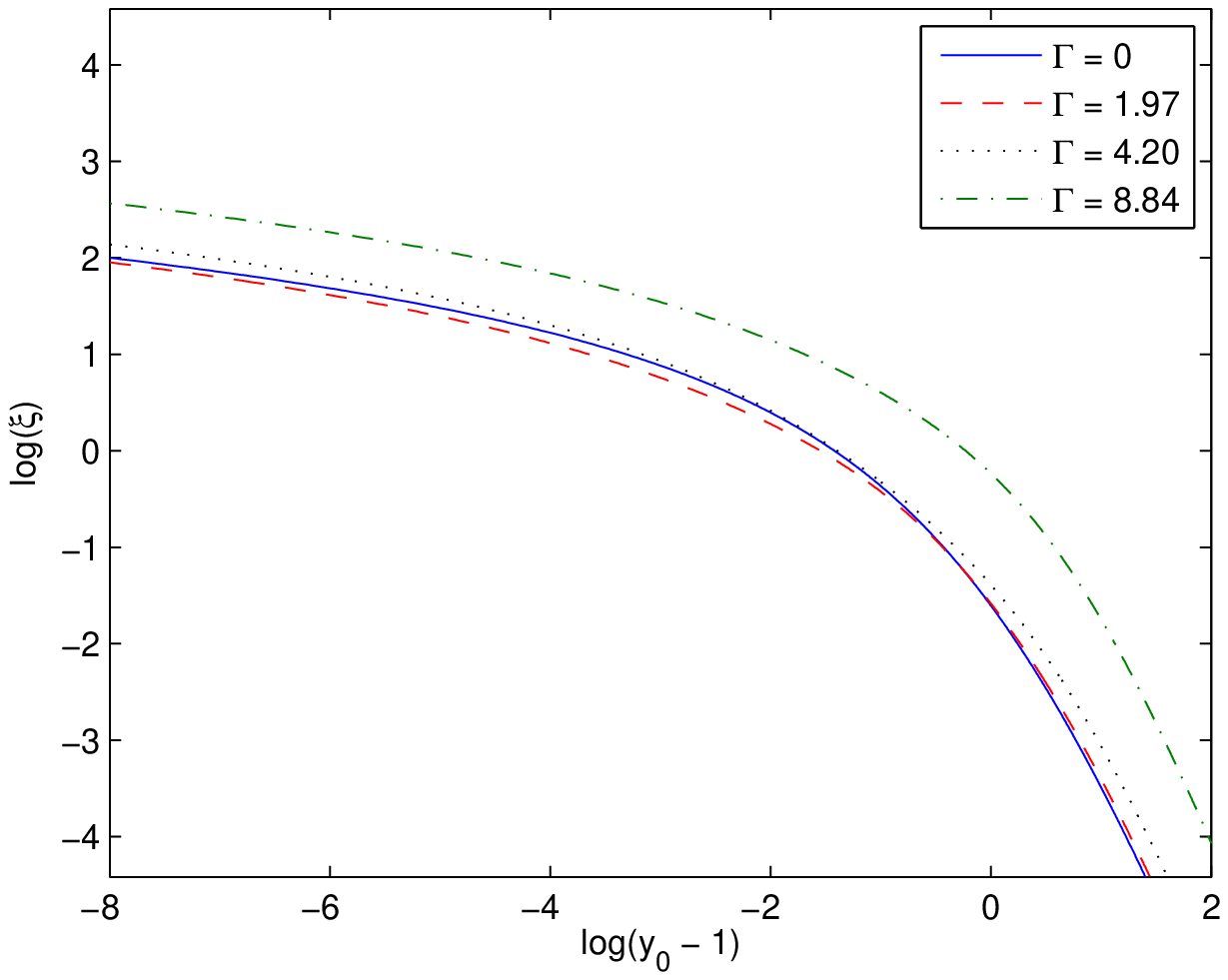}
  \caption{}
\end{subfigure}
}
\begin{subfigure}[b]{0.5\textwidth}
  \includegraphics[width=\textwidth]{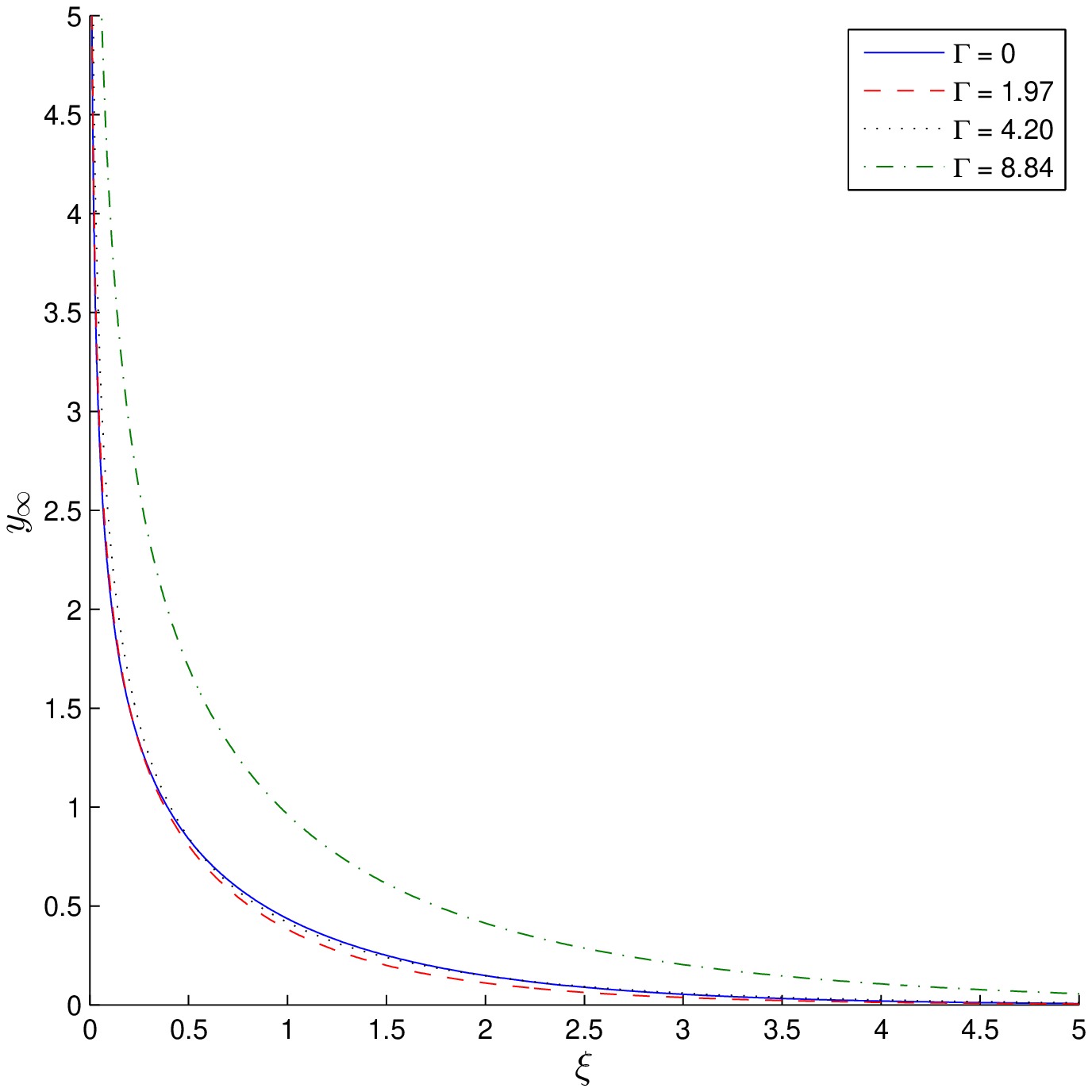}
  \caption{}
\end{subfigure}

\caption{Dependence of drift $\xi$ on (a) the vortex circulation
  $\Gamma$ for initial particle positions $y_0 \in \{
  1.001,1.002,\dots, 1.01,1.02,\dots,1.1,1.2,\dots,2.0\}$ (larger
  $y_0$ corresponding to lower curves), (b) the initial distance $y_0$
  and (c) the distance $y_\infty$ from the flow centerline measured at
  infinity, cf.~\eqref{eq:C}, for the circulation values indicated in
  the legend.}
\label{fig:xi}
\end{figure}

Next, in Figure \ref{fig:xi}a, we show the drift $\xi$ of the
individual particles as a function of the circulation $\Gamma$ and, in
Figure \ref{fig:xi}b, as a function of the initial distance $y_0$ form
the horizontal axis. Since this is how it is often presented, the
latter data is replotted in Figure \ref{fig:xi}c using the linear
scaling with the drift $\xi$ marked on the horizontal axis and the
vertical axis representing $y_{\infty}$, cf.~\eqref{eq:C}.  In Figure
\ref{fig:xi}a we see that the dependence of the drift $\xi$ on the
vortex circulation $\Gamma$ is not monotonous regardless of the
initial position of the particle. Moreover, in a certain range of
$\Gamma$ there are {\em two} initial positions $y_0$ such that the
corresponding drift $\xi(y_0)$ is equal to the drift in the wakeless
flow. While for sufficiently large circulations the drift ultimately
increases as compared to the wakeless flow (corresponding to $\Gamma =
0$), for small values of $\Gamma$ the drift is actually reduced. In
other words, for every $y_0 > 1$ there exists a ``critical''
circulation $\Gamma_0 > 0$ such that the F\"oppl flow has the same
drift $\xi$ as the wakeless flow. This critical circulation is a
nonmonotonous function of the distance $y_0$ from the flow centerline.
In addition to confirming these observations, Figure \ref{fig:xi}b
shows that drift $\xi(y_0)$ is a decreasing function of $y_0$ which
exhibits two distinct asymptotic regimes (see Section
\ref{sec:analysis} for more details on this).

\begin{figure}
\centering
\mbox{
\begin{subfigure}[b]{0.5\textwidth}
  \includegraphics[width=\textwidth]{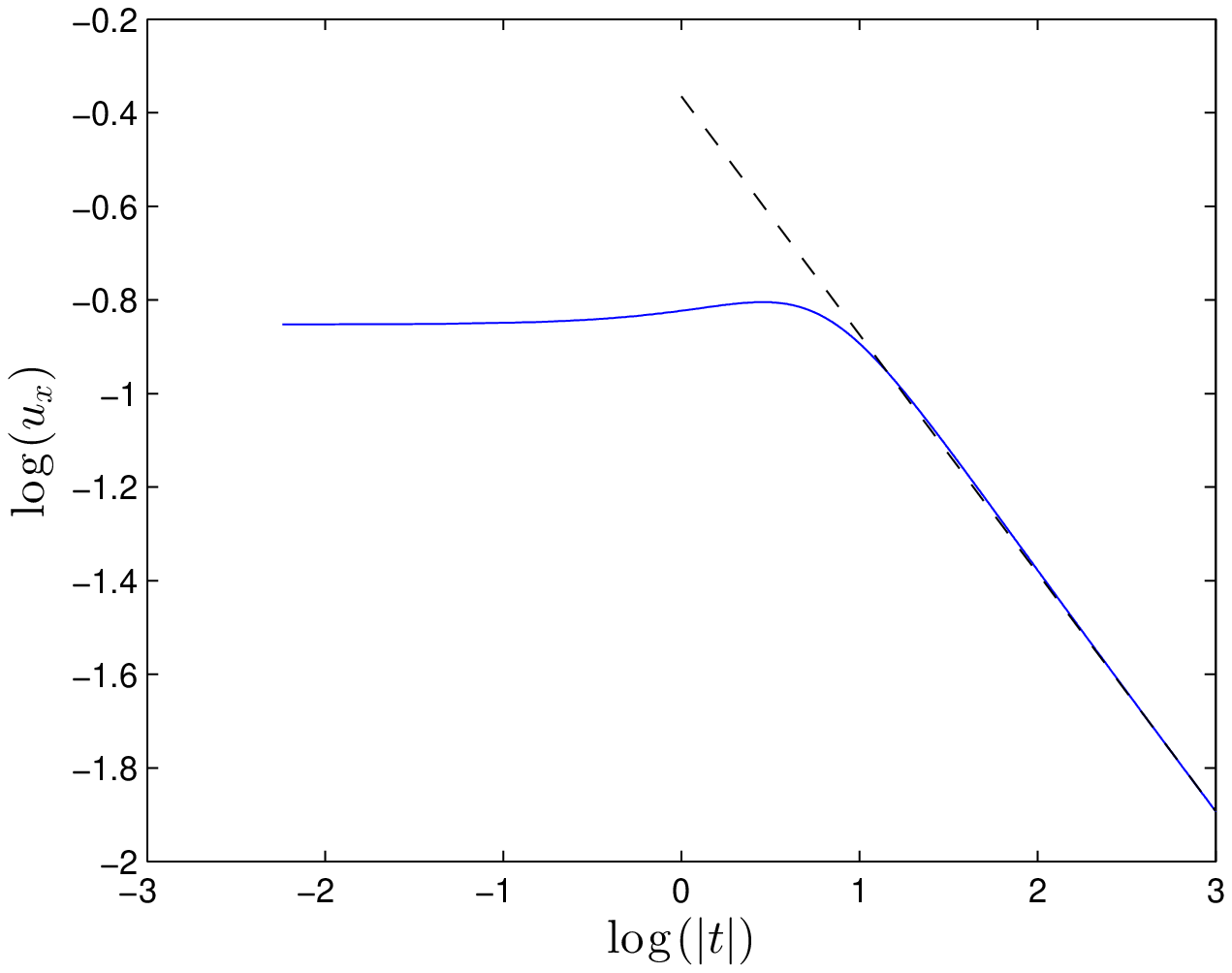}
  \caption{$t < 0$, $\beta = -0.5092$.}
\end{subfigure}
\begin{subfigure}[b]{0.5\textwidth}
  \includegraphics[width=\textwidth]{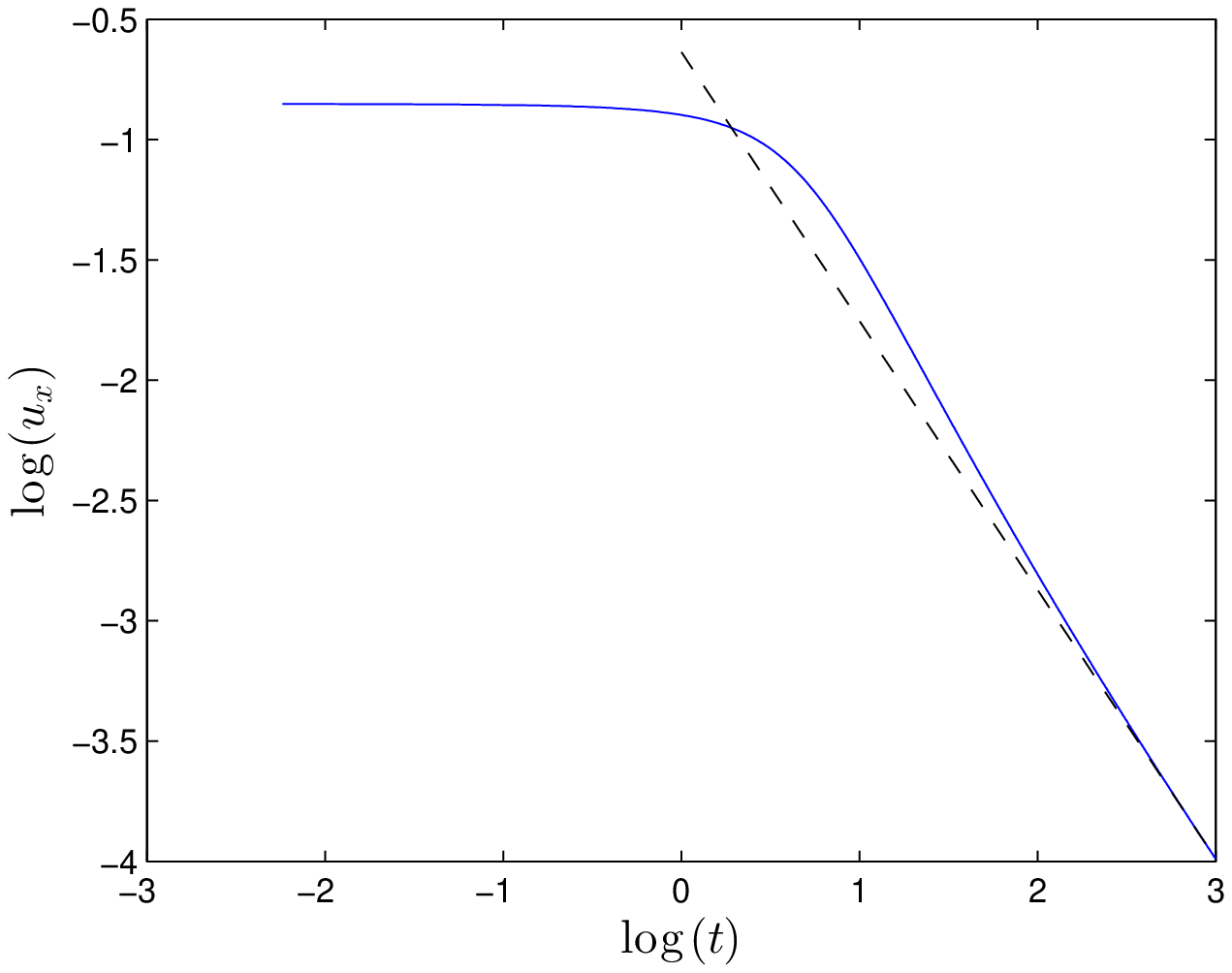}
  \caption{$t > 0$, $\beta = -1.1172$.}
\end{subfigure}
}
\caption{Behavior of the velocity component $u_x$ following the
  trajectory of the particle located at $y_0 = 5$ at $t=0$ in
  Kirchhoff flow for large (a) negative and (b) positive times (solid
  line); the dashed line represents the power-law fit \eqref{eq:h} with
  the exponent values indicated in the captions.}
\label{fig:xi3}
\end{figure}

It turns out that, regardless of the initial position $\x_0 =
[0,y_0]^T$, in the Kirchhoff flow drift \eqref{eq:xi} is {\em
  unbounded}. This is evident from Figure \ref{fig:xi3} showing an
extrapolation using formula \eqref{eq:h} of the velocity component
$u_x(t)$ following the particle trajectory for large positive and
negative times. We observe that, while for positive times the
asymptotic behavior is characterized by the exponent $\beta =
-1.1172$, for negative times the exponent is $\beta = -0.5092$
implying that $u_x(t)$ is not in fact integrable. Although for brevity
in Figure \ref{fig:xi3} the data was shown for one trajectory only
(corresponding to $y_0=5$), analogous results we also obtained for
other trajectories.  Thus, the drift data is not shown for the
Kirchhoff flow in Figure \ref{fig:xi}.

\begin{figure}
\center{\includegraphics[width=0.7\linewidth]{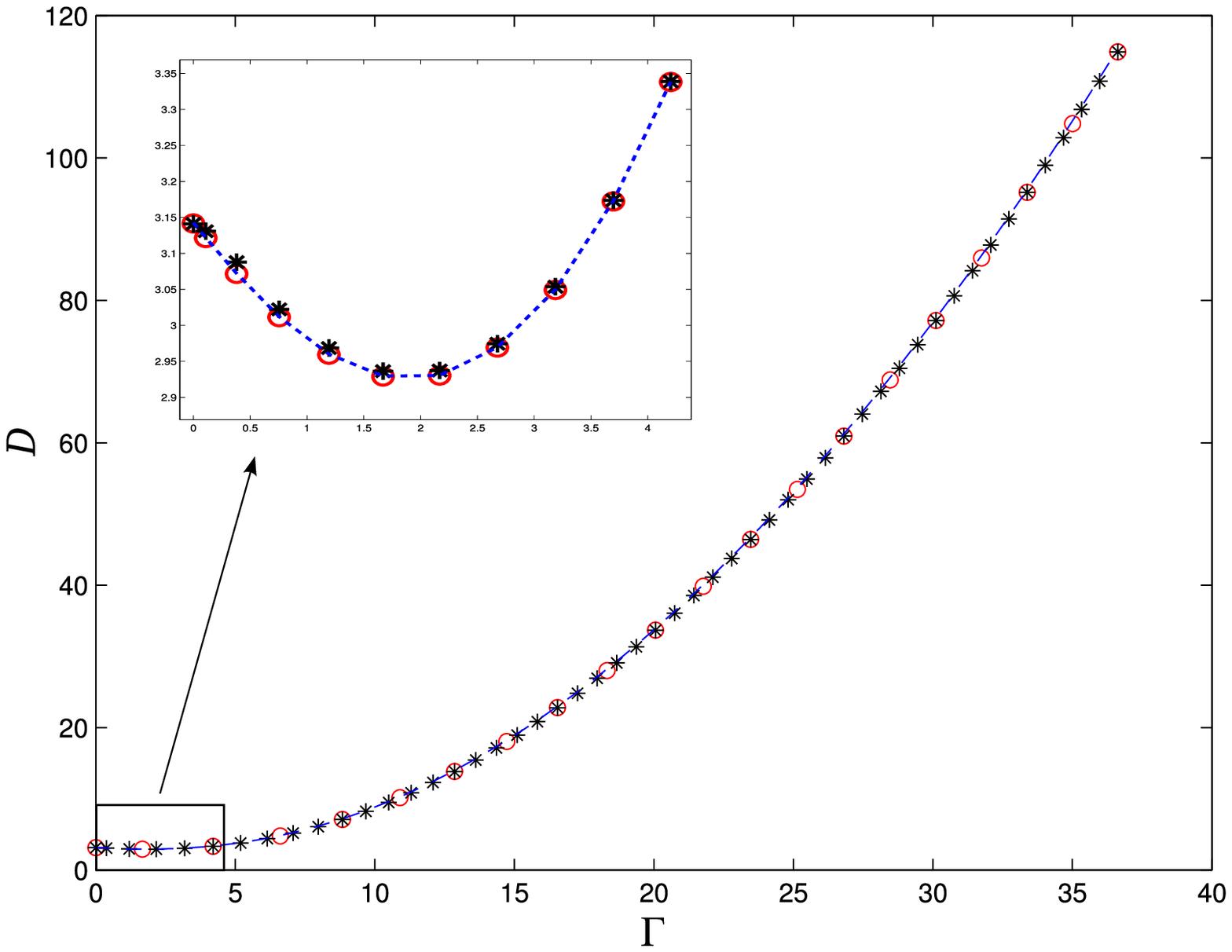}}
\caption{Total drift area $D$ in the F\"oppl flows as a function of
  the vortex circulation $\Gamma$ evaluated based on definition
  formula \eqref{eq:D2} (empty circles), added-mass formula
  \eqref{eq:M} (crosses) and Taylor's theorem
  \eqref{eq:M2}--\eqref{eq:B} (dashed line).}
\label{fig:D}
\end{figure}

Finally, in Figure \ref{fig:D}, we show the dependence of the total
drift area $D$ on the vortex circulation $\Gamma$ computed using the
three methods discussed in Section \ref{sec:compute}, all of which
show excellent agreement. We see that the total drift area exhibits a
well-defined minimum which is a manifestation of the competing effects
observed in Figure \ref{fig:xi}a. The smallest drift area $D = 2.93$
is achieved for $\Gamma = 1.97$, whereas for $\Gamma = 3.6$ drift area
is approximately $D = \pi$, the same as in the wakeless potential
flow. The particle trajectories corresponding to these two cases are
shown in Figures \ref{fig:traj}b and \ref{fig:traj}c.

\section{Asymptotic Analysis}
\label{sec:analysis}

As was discussed in Section \ref{sec:comparison}, the drift of a
particle in F\"oppl flow depends on two parameters, namely, the vortex
circulation $\Gamma$ and the initial distance $y_0$ between the
particle and the flow centerline. In this section we derive
expressions characterizing the drift when the parameters take some
limiting values. The asymptotic study of the drift in the wakeless
potential flow as $y_0 \rightarrow 1$ and $y_0 \rightarrow \infty$ is
presented in \cite{childress2009}, and our approach will build on this
analysis.

A first, trivial, observation is that in the limit $\Gamma \rightarrow
0$ the drift of the wakeless potential flow is obtained uniformly in
$y_0$. Here we will consider the limit $y_0 \rightarrow \infty$.
Since the required transformations are rather complicated, requiring
the use of symbolic algebra tools (Maple), for brevity below we will
only highlight the key steps.

\begin{figure}
\centering
\mbox{
\begin{subfigure}[b]{0.48\textwidth}
  \includegraphics[width=\textwidth]{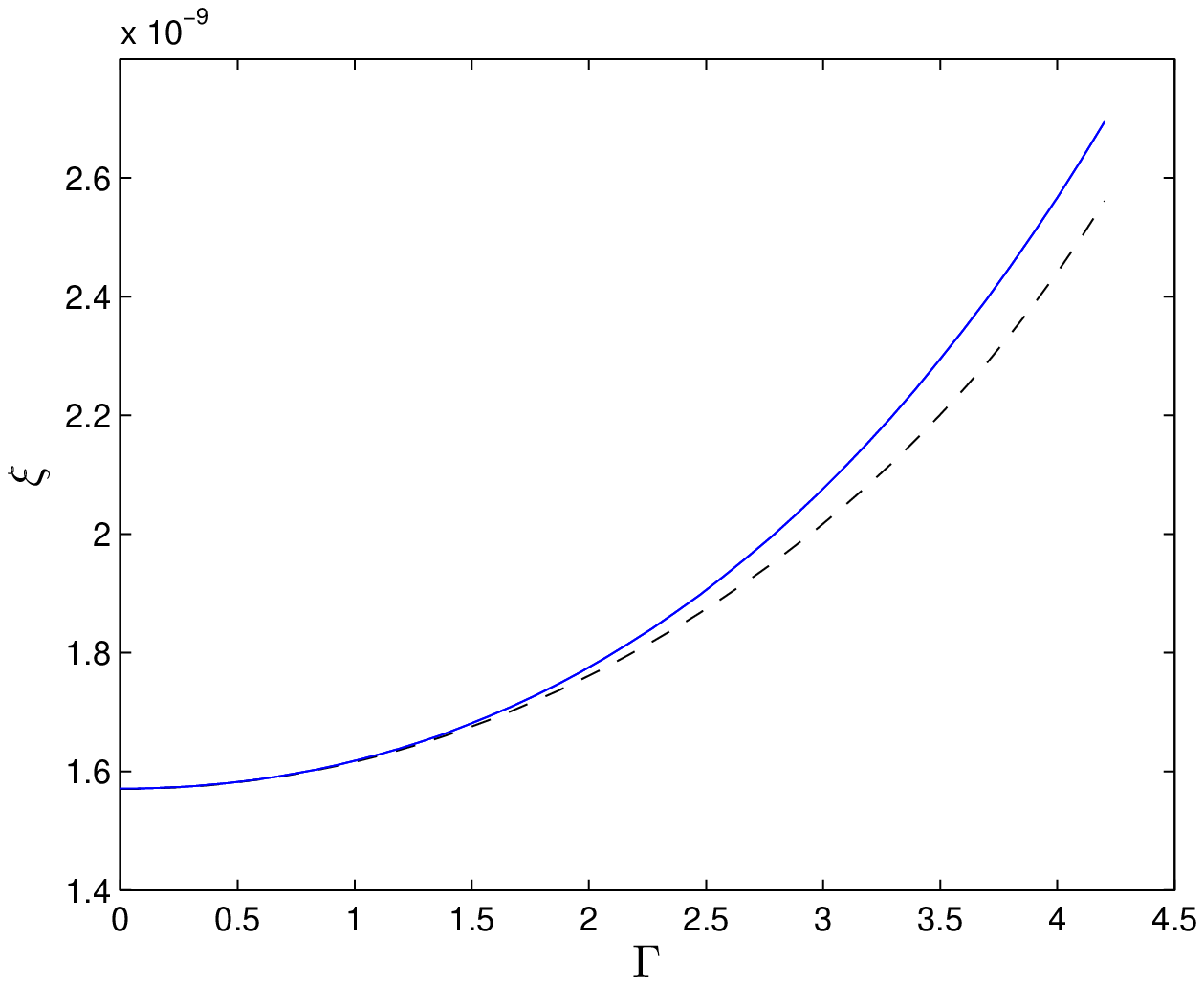}
  \caption{}
\end{subfigure}
\quad
\begin{subfigure}[b]{0.48\textwidth}
  \includegraphics[width=\textwidth]{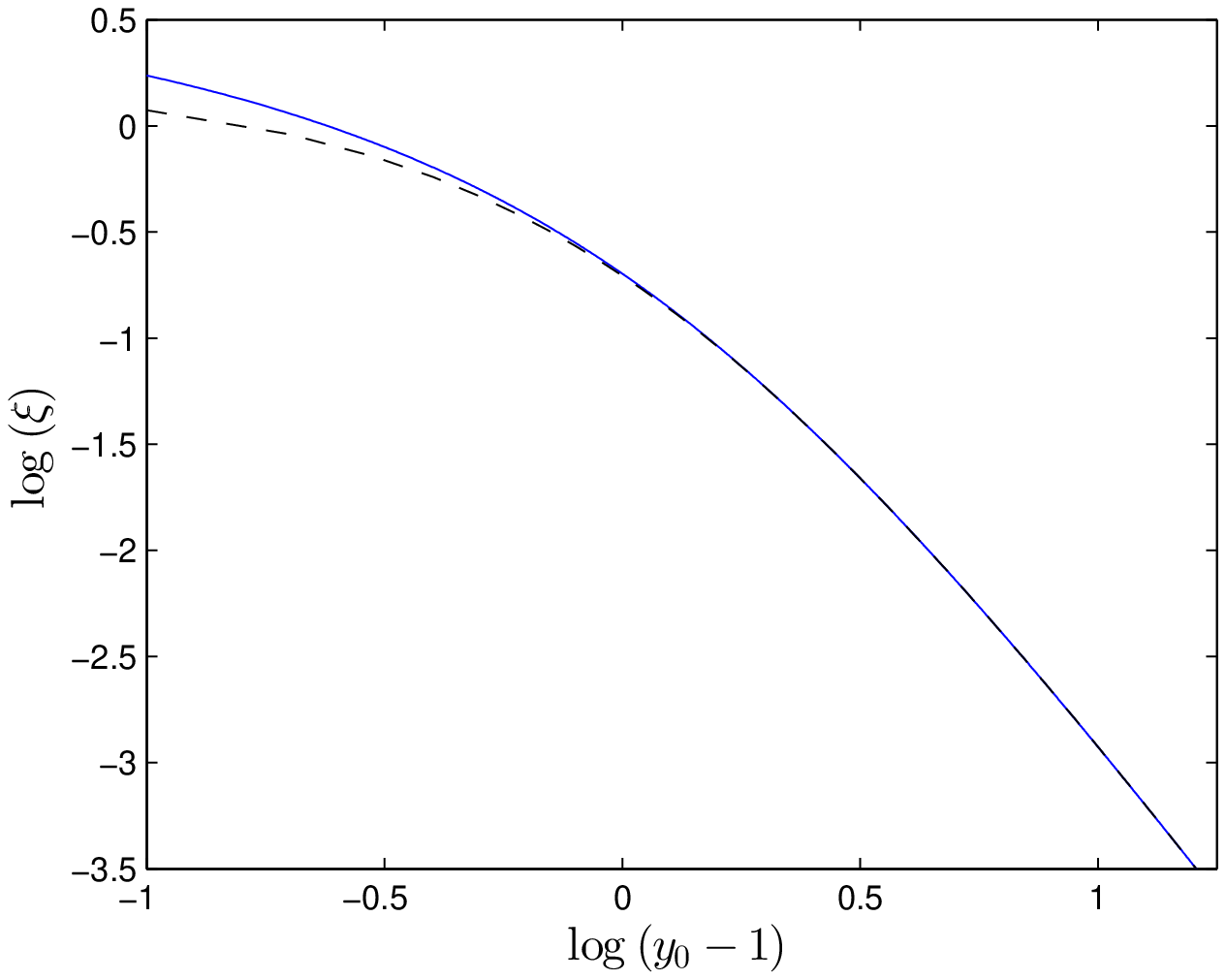}
  \caption{}
\end{subfigure}
}
\caption{Dependence of drift $\xi$ on (a) the circulation $\Gamma$ for
  $y_0=1000$ and (b) the initial particle position $y_0$ for $\Gamma=
  0.38023$; solid lines represent the actual data whereas the dashed
  lines correspond to asymptotic formula \eqref{eq:far_xi}.}
\label{fig:far_xi}
\end{figure}

We start by taking the Taylor expansion of the velocity component
$u_x$ in equation \eqref{eq:dxdt} about the initial position of the
particle $\x_0 = [0, y_0]^T$ and truncate it at the order
$\O(\|\x-\x_0\|^2)$. This is justified by the observation, cf.~Figure
\ref{fig:traj}, that for large $y_0$ the particle trajectories are
close to being circular and in the proximity of $\x_0$ Our goal will
be to integrate this expansion with respect to time,
cf.~\eqref{eq:xi}, but first we have to substitute for $x(t)$ and
$y(t)$ to make the expansion a function of $t$ only. In the limit $y_0
\rightarrow \infty$ the trajectories $\x(t)$ can be approximated with
the solutions $\tilde{\x}(t) = [\tilde{x}(t), \tilde{y}(t) ]^T$ of
system \eqref{eq:dxdt} in which the right-hand side is evaluated at
$\x_0$, i.e., $d\tilde{\x}(t) / dt = \u([0,y_0]^T,t)$, which is
written out as
\begin{subequations}
\label{eq:dtxdt}
\begin{align}  
\frac{d\tilde{x}}{dt}  =& \frac{t^2-y_0^2}{(t^2 + y_0^2)^2} + \frac{\Gamma}{2 \pi} \left[ -\frac{y_0 - y_1}{(-t - x_1)^2 + (y_0 - y_1)^2} + \frac{y_0 - \frac{y_1}{x_1^2 + y_1^2}}{(-t - \frac{x_1}{x_1^2 + y_1^2})^2 + (y_0 - \frac{y_1}{x_1^2 + y_1^2})^2}  \right. \nonumber \\
& \left. + \frac{y_0 + y_1}{(-t - x_1)^2 + (y_0 + y_1)^2} -\frac{y_0 + \frac{y_1}{x_1^2 + y_1^2}}{(-t\frac{x_1}{x_1^2 + y_1^2})^2 + (y_0 + \frac{y_1}{x_1^2 +    y_1^2})^2} \right], \label{eq:dtxdta} \\ 
\frac{d\tilde{y}}{dt} =& - \frac{2 t y_0}{(t^2+y_0^2)^2}	+ \frac{\Gamma}{2 \pi} \left[ \frac{-t - x_1}{(-t - x_1)^2 + (y_0 - y_1)^2} - \frac{-t - \frac{x_1}{x_1^2 + y_1^2}}{(-t - \frac{x_1}{x_1^2 + y_1^2})^2 + (y_0 - \frac{y_1}{x_1^2 + y_1^2})^2} \right. \nonumber \\
& \left.- \frac{-t-x_1}{(-t - x_1)^2 + (y_0 + y_1)^2} + \frac{-t - \frac{x_1}{x_1^2 + y_1^2}}{(-t -      \frac{x_1}{x_1^2 + y_1^2})^2 + (y_0 + \frac{y_1}{x_1^2 +    y_1^2})^2}\right]. \label{eq:dtxdtb}
\end{align}
\end{subequations}
Relations \eqref{eq:dtxdta}--\eqref{eq:dtxdtb} are integrated
analytically for $\tilde{x}(t)$ and $\tilde{y}(t)$ and, before the
resulting expressions are substituted in the series expansion of
$u_x$, they are expanded in a Taylor series with respect to $\Gamma$
which is assumed small. Noting \eqref{eq:G0} and relations $y_1 =
(r_1^2-1) / (2r_1)$ and $r_1 = \sqrt{2\sqrt{x_1^4 - x_1^2 + 1} + 2 x_1^2 -
  1}\, / \,\sqrt{3}$, this expansion can be re-expressed only in terms
of $x_1$, which is the downstream coordinate of the F\"oppl vortex.
Finally, integrating the resulting expression from $t=-\infty$ to
$t=\infty$ and keeping only the leading-order term in $y_0$, we obtain
the following approximation to the drift
\begin{equation} 
\label{eq:far_xi}
\xi = \frac{\pi}{2 y_0^3}
\left[ 1 + 64(x_1 + 1)^4 + 192((x_1 + 1)^5 + (x_1 + 1)^6) \right] + \mathcal{O}((x_1 + 1)^7) 
\end{equation}
valid for $y_0 \rightarrow \infty$ and $x_1 \rightarrow -1$
(equivalently, $\Gamma \rightarrow 0$). As is evident from this
relation, the presence of the F\"oppl vortices introduces a correction
to the expression ${\pi}/(2 y_0^3)$ characterizing the drift in the
wakeless potential flow in the limit $y_0 \rightarrow \infty$
\cite{childress2009}. Asymptotic relation \eqref{eq:far_xi} is
compared to the actual data for $\Gamma \rightarrow 0$ in Figure
\ref{fig:far_xi}a and for $y_0 \rightarrow \infty$ in Figure
\ref{fig:far_xi}b showing a very good agreement in both cases.
Analysis of the drift in the presence of the F\"oppl vortices in the
limit $y_0 \rightarrow 0$ is more complicated and is beyond the scope
of the present study.

\section{Discussion, Conclusions and Outlook} 
\label{sec:final}

In this study we presented a comprehensive analysis, based on careful
numerical computations supported in some regimes by asymptotic
analysis, of the effects of vortex wakes on the Darwinian drift
induced by steadily translating obstacles. We focused on the F\"oppl
and Kirchhoff flows featuring, respectively, a closed and open wake,
which were compared to the wakeless potential flow used as a
reference. We also discussed three different approaches to the
computation of the total drift area, with the method based on Taylor's
theorem leading to a decomposition of $D$ into a ``universal'' part
and a ``flow-specific'' part, in analogy with the decomposition
established in \cite{pushkin2013} for the Stokes flow.

\begin{figure}
\center{\includegraphics[width=0.65\linewidth]{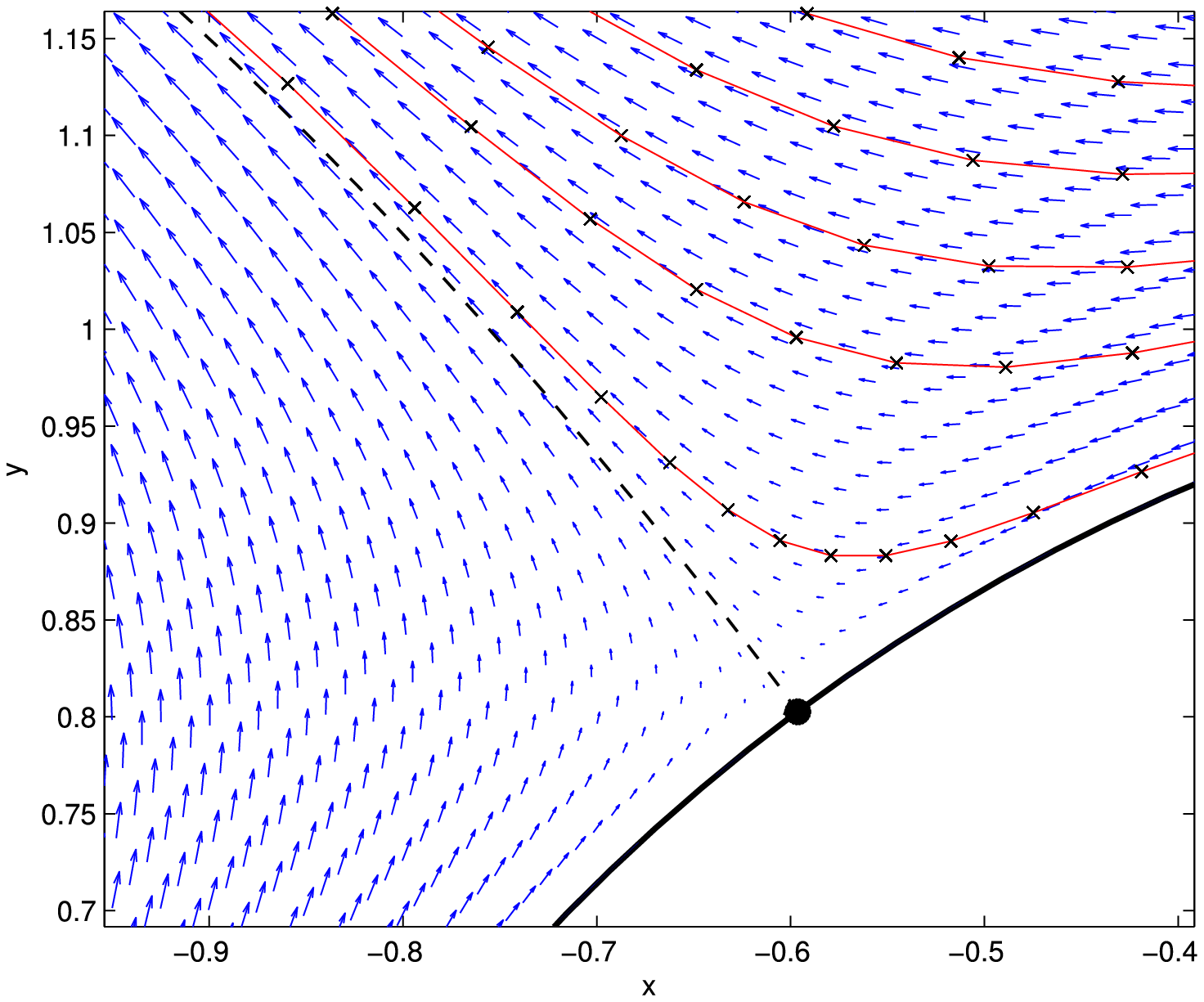}}
\caption{Neighborhood of the stagnation point (marked with a large
  dot) where the recirculation zone separatrix (dashed line) separates
  from the obstacle boundary (thick solid line). Particle trajectories
  are shown with thin solid lines with markers indicating positions at
  equal time intervals.}
\label{fig:stag_point}
\end{figure}

The particle trajectories in F\"oppl and Kirchhoff flows are quite
different (cf.~Figures \ref{fig:traj}b-d and \ref{fig:traj}e). In
F\"oppl flow for certain values of $\Gamma$ and $y_0$ the particle
trajectories exhibit a secondary loop corresponding to the instant of
time when the particle change direction to circumnavigate the
recirculation bubble. An interesting, and perhaps somewhat unexpected,
finding is that while for large values of circulation $\Gamma$ the
presence of the recirculation region in F\"oppl flow increases the
total drift area, an opposite effect occurs for smaller values of
$\Gamma$ (Figure \ref{fig:xi}a). The increase of the total drift area
for large $\Gamma$ can be understood by analyzing the particle
trajectories in the context of changes to the flow topology. Inspection
of Figure \ref{fig:traj}a, corresponding to the wakeless potential
flow, reveals that the largest displacement occurs when the particle
is close to one of the stagnation points (front or rear). The presence
of the wake vortices in F\"oppl flow introduces another stagnation
point (see Figure \ref{fig:traj}b-d) in the neighborhood of which
particles can be trapped and dragged for a long time. This effect is
illustrated in Figure \ref{fig:stag_point} where we show several
particle trajectories in the neighborhood of the separation point
where the boundary of the recirculation zone meets the obstacle.
Symbols on the trajectories mark positions at equal time intervals,
indicating that the particles closer to the separation point are
trapped there for a longer time. Passage near this separation point
corresponds to the climb on the second loop in the trajectories shown
in Figures \ref{fig:traj}b-d. There are some interesting similarities
and differences with respect to the wake effects on the drift in
Stokes flows reported in \cite{pushkin2013}. In both cases the drift
of an individual particle decays as $y_0^{-3}$ when the particle's
position becomes large, cf.~\eqref{eq:far_xi}. This is a consequence
of the fact that both the F\"oppl flow considered here and the
Stokesian swimmer flow studied in \cite{pushkin2013} have a dipolar
far-field representation (even though the spatial dimensions are
different). On the other hand, in contrast to the behavior observed
here, in the Stokes case a significant reflux (negative particle
displacements) was observed resulting in a negative total drift area
corresponding to large wake sizes.

\begin{figure}
\center{\includegraphics[width=0.64\linewidth]{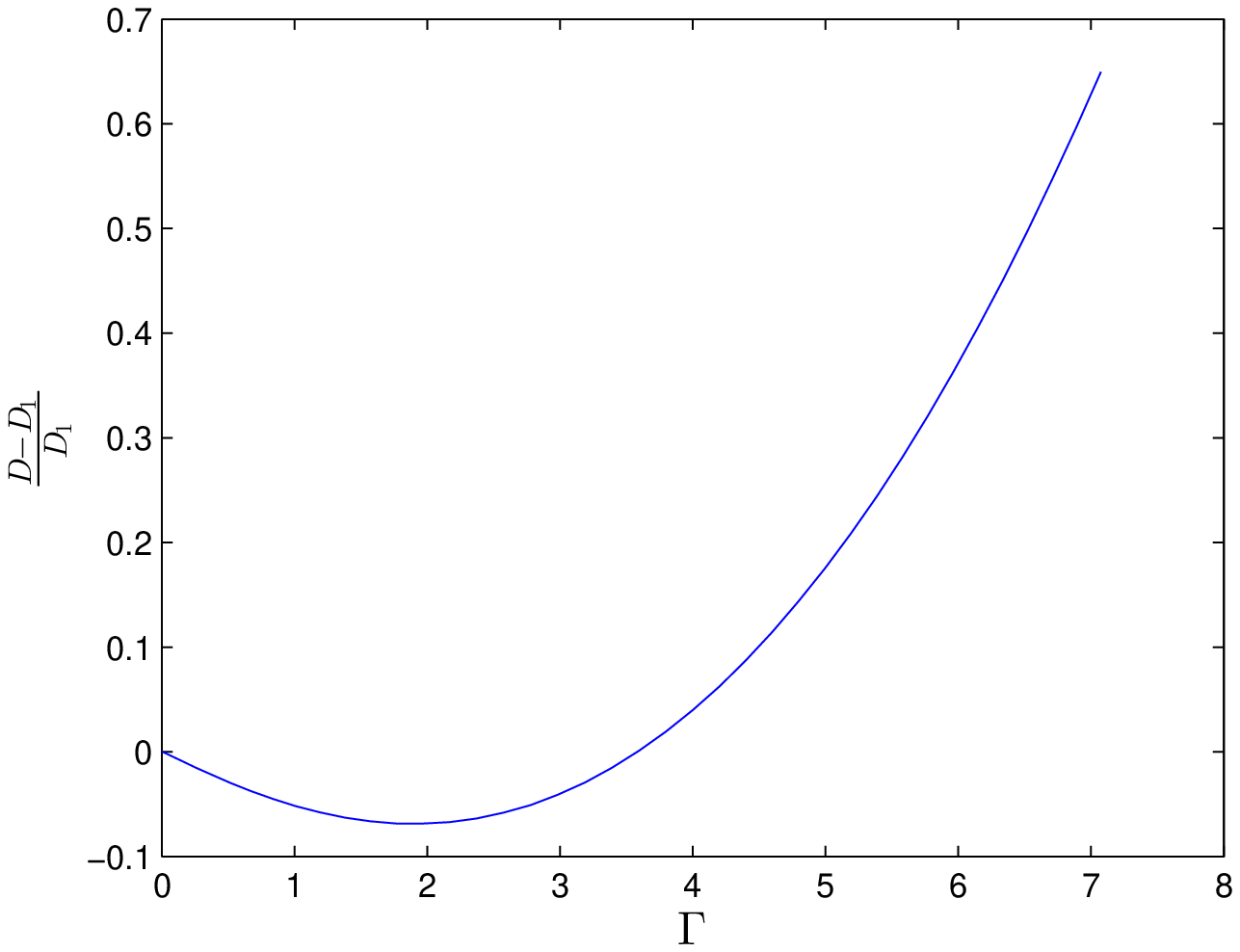}}
\caption{\revt{Relative difference between the drift area in the
    F\"oppl flow ($D$) and in wakeless potential flow ($D_1$) as a
    function of the circulation $\Gamma$.}}
\label{fig:relativediff}
\end{figure}

\revt{We may estimate the drift area using formula \eqref{eq:MM} for
  cases when the size of the wake is smaller than or comparable to the
  size of the cylinder (which is what we may expect in many practical
  situations). If we consider the range $\Gamma \in [0, 7]$, we find
  that the relative difference between the total drift areas in the
  F\"oppl flow and in the wakeless potential flow, i.e., $D$ and
  $D_1$, is approximately $-7$\% when the drift area achieves its
  minimum (see Figure \ref{fig:D}) and $65$\% when $\Gamma \approx 7$.
  In Figure \ref{fig:relativediff} we can see that for the
  representative wakes shown in Figure \ref{fig:traj}b--c, which are
  of relatively small size, $D_1$ may be a good approximation for the
  actual total drift area $D$.  However, for values greater than
  $\Gamma \approx 4.5$, the relative difference exceeds $10$\%.
  Therefore, in practice, when attached vortices are present and are
  large enough, it may be useful to take into consideration their
  effects on the drift area. We add that, as discussed in
  Introduction, this analysis is based on the idealized concept of the
  total drift area and in practical settings, depending on the actual
  travel times of the obstacle, it may be advisable to consider
  partial drift.}

In regard to Kirchhoff flow we demonstrated that drift $\xi$ of
individual particles is in fact not bounded and, consequently, the
total drift area is not defined either. This finding should not be
surprising, given that Kirchhoff flow has an infinite open wake (and
hence can be ``seen'' by the particles as a moving body of an infinite
extent). We note that another instance in which an unbounded total
drift volume was found was the Stokes flow past a spherical droplet
\cite{egd03a}.  Since like Kirchhoff flow and in contrast to the
Stokesian swimmers analyzed in \cite{pushkin2013}, this flow is
characterized by a {\em finite} drag, we may by analogy conjecture
that unbounded total drift area is a feature of steady flows in
unbounded domains which exhibit a nonzero drag.

We expect that the results reported here may help improve the accuracy
of modeling efforts concerning biogenic mixing, such as those
reported in \cite{katija2009}. There is a number of open questions
which may deserve further study concerning, for example, the drift
induced by pairs or larger groups of moving obstacles (in the context
of the potential flow theory, such flows can be studied using the
formalism based on the Schottky-Klein function \cite{c06a}), or
obstacles with asymmetric wakes as were recently reported in
\cite{efz14}. The problem of identifying the shape of the obstacle
which will produce a prescribed drift will lead to some interesting
shape-optimization problems.

\section*{Acknowledgments}

The authors acknowledge the funding provided for this research through
a Discovery Grant of the National Science and Engineering Research
Council (NSERC) of Canada.

\bibliography{mp14a}{}
\bibliographystyle{unsrtnat}

\end{document}